# Aligned metal oxide nanotube arrays: key-aspects of anodic TiO$_2$ nanotube formation and properties

Francesca Riboni[1], Nhat Truong Nguyen[1], Seulgi So[1], Patrik Schmuki[1,2*]

[1]Department of Materials Science WW4-LKO, University of Erlangen-Nuremberg, Martensstrasse 7, 91058 Erlangen, Germany

[2]Department of Chemistry, Faculty of Science, King Abdulaziz University, P.O. Box 80203, Jeddah 21569, Saudi Arabia

*Corresponding author. E-mail: schmuki@ww.uni-erlangen.de

**A B S T R A C T**

Over the past ten years, self-aligned TiO$_2$ nanotubes have attracted tremendous scientific and technological interest due to their anticipated impact on energy conversion, environment remediation and biocompatibility. In the present manuscript, we review fundamental principles that govern the self-organized initiation of anodic TiO$_2$ nanotubes. We start with the fundamental question: Why is self-organization taking place? We illustrate the inherent key mechanistic aspects that lead to tube growth in various different morphologies, such as rippled-walled tubes, smooth tubes, stacks and bamboo-type tubes, and importantly the formation of double-walled TiO$_2$ nanotubes versus single-walled tubes, and the drastic difference in their physical and chemical properties. We show how both double- and single-walled tube layers can be detached from the metallic substrate and exploited for the preparation of robust self-standing membranes. Finally, we show how by selecting the "right" growth approach to TiO$_2$ nanotubes specific functional features can be significantly improved, *e.g.*, an enhanced electron mobility, intrinsic doping, or crystallization into pure anatase at extremely high temperatures can be achieved. This in turn can be exploited in constructing high performance devices based on anodic TiO$_2$ in a wide range of applications.

*Keywords:* self-organizing electrochemical anodization; TiO$_2$ nanotubes; double- and single-walled tubes; TiO$_2$ nanotube robust self-standing membranes; electronic and optical properties.

1. **Introduction**

One-dimensional (1D) nano-architectures exhibit a unique set of properties that attracts a great deal of scientific and technological interest. In particular, 1D-nanotube (NT) arrays have gained increasingly more attention as they provide an exceptional combination of optical, electrical and chemical properties with extreme geometry that has inspired remarkable advances in recent nanotechnology and microelectronics. Compared to bulk materials, NTs can feature characteristics such as high electron mobility and quantum size related effects, along with very high specific surface area and mechanical strength that are highly relevant for numerous applications in physics, chemistry, materials science, and medicine.[1–3]

Upon the discovery of carbon nanotubes,[4] and following their large impact on nanotechnology,[5–11] research efforts have also been extended to a large number of inorganic materials, mainly oxides and sulfides of metals and transition metals.[12–18] For oxide/sulfide nanotubes, a variety of synthesis routes has been reported such as hydrothermal,[19–21] sol-gel or template-assisted methods.[22–25] In recent years, another most frequently adopted and straightforward approach for growing highly ordered 1D-nanostructures is self-organizing electrochemical anodization (SOA).[26,27] It allows the fabrication of vertically oriented, size-controlled and back-contacted (*i.e.*, anchored to a metallic substrate) nanostructured arrays (*e.g.*, aligned pores, nanochannels, and nanotubes) on various metals and alloys (Fig. 2).[28–36] As illustrated in Fig. 1a, anodizing can be carried out in a simple anode/cathode arrangement, where the metal of interest (M) serves as anode. The oxide growth process is based on the oxidation of M (M → $M^{z+}$ + $ze^-$) and its conversion to metal oxide ($MO_{z/2}$) under a suitable voltage, the source of oxygen ions being typically $H_2O$ in the electrolyte. Depending on the electrolyte and on several anodization conditions, three different scenarios are possible: (i) if the oxide is completely soluble ($M^{z+}$ ions are entirely solvatized) in the electrolyte, the layer is continuously etched and metal corrosion (or electropolishing, EP) is observed (Fig. 1a, EP); (ii) if the formed oxide is entirely insoluble in the electrolyte, a compact oxide (CO) layer is formed on the surface of the metal (Fig. 1a, CO); (iii) if the oxide is partially soluble (cations can be solvatized) and under a specific set of experimental conditions, a steady-state between oxide formation and dissolution can be established and a porous or tubular metal oxide can grow (Fig. 1a, PO).

Historically, landmarks on the growth and use of such electrochemical nanostructures were certainly represented by the initial reports of Rummel[37] and Baumann[38] on the fabrication of porous-type aluminum oxide layers, and by the work of Thompson and Wood[39] who first highlighted the great potential of porous alumina for organic dyes or inorganic pigment loading as well as corrosion protection. A cornerstone towards functional uses of porous alumina was the report of Masuda *et al.* in 1995 that demonstrated a virtually perfect hexagonal arrangement of closely-packed cells if porous alumina is grown under the "right" conditions in acidic electrolytes.[40] Over the last 20 years, porous alumina has been extensively applied as a template for the synthesis of functional nanomaterials, particularly for the fabrication of nanorods, nanowires and nanotubes by deposition of metals, semiconductors or polymer materials.[40–44] In addition, also the direct use of nanoporous $Al_2O_3$ for decorative, wear- and corrosion-resistant applications, as a filter or as a photonic crystal, has been strongly developed.[45] Over decades, aluminum was considered the only metal that can form anodic self-ordered oxide structures and early works by Assefpour-Dezfuly[46] and Zwilling[47] on $TiO_2$ nanotube formation on Ti went almost unnoticed. A few years later, this type of first generation of $TiO_2$ nanotubes grown in acidic aqueous electrolytes was further investigated and essentially confirmed the earlier findings.[48,49] However, all these layers were not highly organized, limited in length to ~500 nm, and with considerable wall inhomogeneity, but anyway proved that the key-factor for growing self-organized $TiO_2$ nanotubular arrays is the addition of fluoride species to the electrolyte.

Reasoning for this specific role of fluorides is that they provide the feasibility to establish an equilibrium between anodic oxide formation and chemical dissolution ($Ti^{4+}$ is relatively easily solvatized by $F^-$ in the form of $[TiF_6^{2-}]$). Further significant improvements were the introduction of electrolyte pH mediation,[50] the use of non-aqueous fluoride electrolytes (glycerol, ethylene glycol, DMSO),[51–54] and fine tuning of voltage control and alteration procedures,[55–57] that allowed a higher degree of self-ordering and a strongly enhanced morphology control. Progressively, the use of fluoride-based electrolytes was shown to enable the growth of similar nanotubular anodic oxides also on a wide palette of metals, such as Fe,[58–60] W,[61,62] Nb,[63,64] V,[65] Co,[66] Ta,[67–70] Zr,[71–73] Hf[74,75] as illustrated in Fig. 2 and even metal-alloys, such as Ti-Nb,[28,76–81] Ti-Zr,[82] Ti-Ta,[33,83–85] Ti-W,[36] Ti-Cr,[86] Ti-Mo,[87] Ti-V,[77] Ti-Ru.[88] In particular, anodizing metal-alloys turned out to be a unique and elegant way of direct doping the tube oxide layer by introducing well-defined amounts of a secondary species in the oxide, or to intrinsically decorate tubes with noble metal nanoparticles.[89,90]

Owing to the extended control over nanoscale geometry, direct back-contacting, and due to the large versatility of this approach to grow nanotubular structures of various metal oxides and mixed oxide compounds, electrochemical anodization is in many cases a most straightforward nanotube synthesis path. Therefore, self-organized nanotubular structures find application in a variety of fields: $TiO_2$ nanotubes are extensively investigated in solar cells (DSSCs), photocatalysis and biomedical applications, α-$Fe_2O_3$ NTs as anode for (photo)electrochemical water splitting reactions,[59,60,91,92] ordered nanoporous $V_2O_5$ and $Co_3O_4$ structures for batteries and ion intercalation devices or as oxygen evolution catalysts,[65,66] $WO_3$ nanochannels for gas sensing and electrochromic applications,[93–95] etc.

However, $TiO_2$ represents by far the most investigated metal oxide.[27,49,59,64,65,95,96] This can be ascribed to the outstanding breadth of properties of this compound: classically used for its non-toxic, environmentally friendly, corrosion-resistant and biocompatible characteristics, many functional applications rely on the specific ionic and electronic properties of $TiO_2$. Among these, a wide range of photo-electrochemical applications (photocatalytic splitting of water into oxygen and hydrogen, self-cleaning surfaces, organic pollutant degradation) or the use of $TiO_2$ in solar cells (most frequently in Grätzel-type DSSCs) are most common.[97,98] Aligned one-dimensional arrangements are particularly beneficial for the performance in (photo)electrochemical applications, as orthogonal carrier separation in these structures is facilitated, *i.e.*, electrons and holes are driven apart (that is, $e^-$ are collected at the back contact and $h^+$ accumulated at the semiconductor/electrolyte interface). This particularly helps overcoming the limitations of a short diffusion length of holes in $TiO_2$ (~10 nm) while exploiting the comparably long electron diffusion length (~20 μm in $TiO_2$ nanotubes).[99] As $TiO_2$ nanotubular structures simultaneously combine the highly functional features of $TiO_2$ with a regular and controllable nanoscale geometry (that is, tube-length, -diameter, and self-ordering that can be tailored over large length scales), intense research activities have been devoted to anodic $TiO_2$ nanotubes over the last ten years (see, for instance, research trends reported in Refs. 27,100).

While a number of recent reviews focus on properties and applications of $TiO_2$ nanotubes and other aligned channels,[26,27,101] the present review mainly focuses on mechanistic aspects of nanotube growth and self-organization: why do tubes (not pores) form, what are the critical factors to achieve self-organization and how to improve it, why is the tube wall morphology sometimes rippled sometimes smooth, how to grow multi-layered tubes, why are single-walled $TiO_2$ nanotubes so superior compared to double-walled tubes? Along with the factors determining nanotube formation and leading to various tubes morphologies, we will also briefly discuss the significance of various types of tubes and tube features for electric, electronic and optical properties that in turn dramatically affect the performance of tube arrays in virtually any (photo)electrochemical application.

## 2. Stages and key aspects of nanotube formation

As mentioned, anodization is frequently carried out in a 2-electrode electrochemical arrangement as illustrated in Fig. 1a, with the metal of interest (M) as working electrode (anode) and a counter electrode (typically platinum or carbon) as cathode. On various metals (valve metals), oxidation (1) and conversion to a thick metal oxide (2) can be achieved in an aqueous electrolyte at sufficiently high anodic voltages. For non-valve metals, only thin oxides can grow, as $O_2$ evolution (3) is the preferred anodic reaction.[26,101,102] In either case, a counter reaction, typically the reduction of $H^+$ (from the electrolyte) to gaseous $H_2$ (4), takes place at the cathode:[26,27]

$$M \rightarrow M^{z+} + z\, e^- \tag{1}$$

$$M + \frac{z}{2} H_2O \rightarrow MO_{z/2} + z\, H^+ + z\, e^- \tag{2}$$

$$H_2O \rightarrow O_2(g) + 4H^+ + 4e^- \tag{3}$$

$$z\, H_2O + z\, e^- \rightarrow \frac{z}{2} H_2\uparrow + z\, OH^- \tag{4}$$

Mechanistically, anodic oxide formation is controlled by a high-field ion formation/transport process, where the kinetics of oxide growth (at the metal/oxide or at the oxide/solution interface) depends to a large extent on the rate of ion migration where $M^{z+}$ migrates with the field outward and $O^{2-}$ (abstracted from water) inward (see Fig. 1b). Ion movement is influenced by the relative mobility of the cation and the anion in the oxide. Accordingly, the film can grow at the inner and outer interface.[103] Generally, the overall oxide growth is related to the electric field (E = ΔU/d) applied across the oxide layer:

$$I = A\, \exp(BE) = A\, \exp(B\Delta U/d) \tag{5}$$

where I is the current, $\Delta U$ is the voltage across the oxide layer of thickness $d$, and A and B are material-dependent constants.[104] If an insoluble compact oxide is formed, the anodization process is self-limiting: *i.e.*, for a given voltage ($\Delta U$) and with an increasing thickening of the oxide layer ($d$), the field E drops over time to such an extent that significant ion migration cannot be promoted any longer and a compact oxide (CO) layer with finite thickness $d = f \times U$, where $f$ is film formation factor ($f(TiO_2)$ ~2.5 nm/V)[105], forms (Fig. 1a, CO).[106–108] By contrast, if all $M^{z+}$ ions are solvatized, for instance, in the form of water soluble metal-fluoro complexes $[MF_x]^{(z-x)-}$ in fluoride-based electrolytes (6):

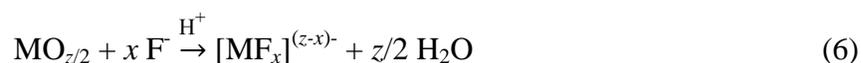

$$MO_{z/2} + x\,F^- \xrightarrow{H^+} [MF_x]^{(z-x)-} + z/2\,H_2O \tag{6}$$

and dissolved, no oxide film is grown and active corrosion or electropolishing (EP) of the metal is observed (Fig. 1a, EP). Only if an equilibrium between oxide formation (2) and dissolution (6) is established, oxide growth can continue and form 3D structures, *i.e.*, under a specific set of conditions, such as temperature, applied potential, aqueous or organic electrolyte, pH and specific ions, self-organized anodic nanostructures can be achieved (Fig. 1a, PO).

A versatile tool to describe thermodynamics and kinetics of electrochemical reactions are current-voltage curves (where the potential can be regarded as thermodynamic axis and the current as kinetic axis).[109] To monitor the evolution of nanotubes over time, it is interesting to follow $j$-t curves. A typical current-time curve (Fig. 1c) that describes the conditions to nanotube formation can be divided into three sections: (i) in the early stage of anodizing, a compact oxide layer is formed; as the barrier layer thickens, $j$ accordingly decreases (this, in line with (5)); (ii) under the "right" conditions (see below), irregular nanoscale pores penetrate into the initial compact oxide; here, the current rises as the reactive area (exposed to the electrolyte) increases; (iii) a regular nanoporous/nanotubular layer forms and the current drops again until a steady-state situation is established between oxide formation at the metal/oxide interface and the dissolution at the oxide/solution interface ($M^{z+}$ ions from dissolved oxide and ions directly ejected from the oxide by the high field are solvatized by $F^-$, (6)). Due to this dissolution, a current significantly higher than that corresponding to compact oxide formation (assuming the same applied voltage) is observed. According to (5), this means that under steady-state condition for tube growth a higher field is present at the tube bottom (that drives ion migration).

However, except for this steady-state formation-dissolution equilibrium, a second key factor to tube formation is the buildup of compressive stress during the early stages of anodic oxide formation. As predicted by the Pilling–Bedworth (PB) ratio ($\Phi$, the ratio of oxide volume created by oxidation *vs.* the metal volume consumed),[110] stress is generated by volume expansion (for many metal/oxide systems) when converting the metal to oxide and is due to the constraint of volume conservation at the metal/oxide interface. This stress is present in any anodic film but, if a compact layer is formed, the stress remains "frozen in" or leads to some cracks in the oxide layer. In situ measurements of internal stress evolution (Fig. 1d) reveal that two different characteristic stages are observed when anodizing a metal substrate: in the first stage, a relatively large compressive ($d(\sigma_{ox} h_{ox})/dt < 0$) internal stress, in the range of 2–4 GPa for anodic $TiO_2$, is observed that has been, except for the PB effect, also attributed to oxygen evolution (OER) from the electrolyte that occurs inside the oxide film. Here the anodization efficiency ($\varepsilon$, that corresponds to the fraction of oxidized metal ions retained in the final anodic film relative to the film dissolution)[110] is comparably low (*i.e.*, low $dV/dt$ and $\varepsilon \sim 10$–$20\%$) as the majority of charge is consumed in the anodic side reaction (OER). The presence of an overall residual compressive stress in the tube layers is most evident from the instantaneous roll-up of thin NT layers when removed from the substrate (inset of Fig. 1d). When the anodic film thickness reaches a certain value (critical thickness, $T_{cr}$), the transport of electrolyte through the growing oxide becomes too slow and OER ceases due to kinetic limitations. At this stage, also the compressive stress disappears and in some cases, an instantaneous tensile stress component is observed. (*i.e.*, to maintain a compressive stress situation, the oxide should not grow thicker than $T_{cr}$).[111] Or, in other words, only if a steady-state "formation-dissolution equilibrium" is established, compressive stress can permanently act on the continuously formed oxide and finally, in a flow mechanism, shift the oxide up to form tube walls as described in more detail below. Also, it should be noted that during oxide growth a second component of stress, *i.e.* electrostrictive stress, is acting on the anodic system. The extent of this is strongly material- and voltage-dependent.[112]

## 2.1 Pore initiation and nanopattern selection

As pointed out above, during self-ordering electrochemical anodization essentially four concomitant reactions occur: *i)* formation of metal ions upon metal substrate oxidation (1), *ii)* generation of $O^{2-}$ from the reaction of $H_2O$ at the film surface (3), *iii)* dissolution and loss of metal ions to the electrolyte, driven by field-assisted cation ejection (5), and *iv)* chemical dissolution of the film (6). Arguments *i)* and *ii)* are responsible for forming the film, while *iii)* and *iv)* lead to oxide loss.

Experimentally, in the very early stage of anodization, first the formation of a few nanometer-thick conformal oxide layer (barrier or compact layer) takes place. Then, morphological instabilities occur, such as a corrugated surface (Fig. 3a). This process leads to a stochastic distribution of sites with "thinned" oxide locations – this surface pattern is in a lateral length scale of some nm to some 10 nm.[105,108] This causes electrochemical processes to occur accelerated at the thinner parts of the layer, that is, in the "valleys" (Fig. 3a) where the intensity of the applied electrochemical field is enhanced (5). Noteworthy, at these most active sites also the generation of $H^+$ is enhanced ($Ti^{4+} \xrightarrow{H_2O} TiO_x + H^+$; $Ti^{4+} \xrightarrow{HF} TiF_x + H^+$), which may additionally affect the local chemical dissolution (thinning) of the oxide (6).[31,105,113]

Linear stability analysis, based on perturbation analysis[114,115] or numerical treatments,[110] indicate that compressive stress in a thick anodic oxide (compact oxide) is rather small and not sustained by a certain flux of ions (electromigration) across the film/electrolyte interface (*i.e.*, formation-dissolution equilibrium). Therefore, only for thin films, conditions are provided for stress evolution sufficient for anodic pattern formation. As outlined above, this is also dictated by oxide dissolution at the oxide/electrolyte interface. Time-evolution of this interface is a function of electric potential, $\phi$ (*i.e.*, the driving force for metal-ion transport), and of oxygen chemical potential, $\mu_{OX}$ (or driving force for oxygen-ion transport). Both $\phi$ and $\mu_{OX}$ are uniform at the oxide/solution interface, that is, $\phi$ is set to zero as all the small potential drops through the interfacial electrical double layer do not affect the interface dynamics, and $\mu_{OX}$ is also zero as the oxide surface is assumed stress-free; therefore, $\phi$ and $\mu_{OX}$ gradients within the film are distorted. In particular, their gradient is maximized at valleys and attenuated at ridges (Fig. 3a). Correspondingly, both metal and oxygen ions migrate more rapidly at valleys than at ridges. However, given a distinct volume change when a specific metal is converted to oxide (corresponding to a certain Pilling-Bedworth ratio $\Phi$), only a limited set of parameters, and thus of efficiency $\varepsilon$, applies for pattern selection and formation of nanoporous film. For instance, in the case of $TiO_2$, $\varepsilon = 0.50–0.65$ was determined under various experimental conditions for nanotube growth,[116] which is in good agreement with theoretical predictions.[110] According to this model, if the efficiency lies outside the predicted range, no ordered nanotubes would be obtained. In particular, at lower efficiencies the fraction of oxidized metal atoms retained in the oxide is small and, therefore, the dynamics of the oxide/solution interface is dominated by oxygen migration that leads to valleys recession and deepening, destabilizing the interface (*i.e.*, extensive etching of the oxide and, ultimately, corrosion and/or electropolishing are observed). By contrast, at higher efficiencies the interface evolution is regulated by the deposition rate of new oxide and the upper limit is represented by ε approaching 100%, where a compact (thickness-limited) anodic film is formed as any depressions on the oxide/solution interface instantly fill with new material. Thus, nanoporous pattern selection (*i.e.,* an optimal ε range) is governed by a combination of oxygen flow ($\mu_{OX}$), responsible for destabilizing the interface, and of metal ions flow ($\phi$) that promotes stability.[110,116,117]

## 2.2 Growth of nanostructured oxide

Further steady growth of nanotubes is in literature explained by two common models: *i*) either assuming an enhanced field-assisted dissolution of titania along with a field-assisted ejection of $Ti^{4+}$ ions at the base of the pores,[118–120] or *ii*) a field-assisted viscous flow of film material from the barrier layer towards the cell wall regions.[103,121–123] In either case, field enhancement at the curved bottom oxide (Fig. 3d) or flow, self-organization can be achieved only for an optimized set of parameters (*e.g.*, applied potential, current density, temperature and fluoride concentration)[124] which in turn reflect that a set of conditions needs to be established to adjust an optimal oxide formation-dissolution equilibrium. This is consistent with the frequent observation of a barrier layer that is thinner than an equivalent compact oxide under the same applied voltage conditions. The final thickness of compact anodic oxide on Ti is usually $d = f \times U$.[105] In other words, the thinner oxide at the tube bottom reflects that a high field is present that cannot be lowered by oxide growth as all the corresponding dissolution rate prevents this. In pure field driven models, such as established by Gösele *et al.*,[125] the enhanced field (see distribution in Fig. 3d) promotes ejection of $Ti^{4+}$ species to the electrolyte and this would be a sufficient condition for continuous growth. However, recently established models estimate that the forces in the growing oxide are sufficiently high to cause viscous flow of the oxide. While a number of arguments such as the introduction and distribution of incorporated electrolyte ions can be explained by both approaches, there are some interesting findings supporting the presence of a field-assisted flow:

(i) An experiment carried out using a thin W-tracer layer in an Al substrate and then growing $Al_2O_3$ nanopores through this[123,126] (similarly for $TiO_2$ nanotubes)[103] showed a flow-type distribution of the tungsten metal oxide tracer from the barrier layer towards and up the cell walls (Fig. 3b).

(ii) Some reports present data that, at this stage of the tube growth, cation ejection to the electrolyte is negligible[127] and inward migration of anion species mainly contributes to the ionic current in the high-field barrier oxide. Therefore, the formed oxide must be pushed out of the way, *i.e.* up the wall.

(iii) The nanotube or nanopore film thickness (the tube length) exceeds the expected thickness of a corresponding compact oxide. For anodic $TiO_2$, the thickness of a compact film is 2.43 times that of the consumed titanium (corresponding to the Pilling-Bedworth ratio (PBR)). The PBR can be measured or calculated assuming a certain oxide density.[103] If the oxide growth would be ruled by a high field oxidation process only, then the tube or pore length would be 2.43 times the amount of metal consumed. Nevertheless, for Al and Ti the corresponding tubes or pores are longer. (According to models that are restricted to the high field oxidation, an explanation of this finding requires an assumption of an oxide density that is lower than usually reported for amorphous anatase or rutile material).[103] In flow models, this length extension is ascribed to a stress-induced viscous flow of oxide vertically to the metallic film. In other words, $TiO_2$ nanotube growth occurs, at least partially, by a flow of $TiO_2$ up the wall (any lateral expansion of the oxide is negligible).[128] This is illustrated in Fig. 3d where broken and solid lines represent the position of the atoms before and after oxidation, respectively. As the reaction proceeds (stages 1–4), increasingly larger stress is generated in the oxide (*i.e.*, solid lines bend to the vertical axis) and the bent location shifts to the center between two adjacent tubes.[105,122,128,129] Measured expansion factors, $F_m$ (estimated as the thickness of the film relative to that of the oxidized metal) can be expressed as a function of the total volume of the film (P), influenced by incorporated contaminants ($k$):

$$F_m = \frac{k\varepsilon\Phi}{(1-P)} \quad (7)$$

where $\varepsilon$ is the oxide growth efficiency and $\Phi$ the Pilling-Bedworth ratio. Experimental investigations show that $F_m$ for $TiO_2$ nanotubes grown under self-organizing conditions may vary over a relatively wide range (~ 1.4 – 3.0) and strongly depends on experimental conditions, such as the anodization potential and the water content in the electrolyte, while the fluoride content is not of significant influence.[103,122,128] This may indicate that the onset (or occurrence) of flow is parameter-dependent.

## 2.3    Nanopores *vs*. nanotubes

A main difference in porous alumina and $TiO_2$ nanotubes is that, for $TiO_2$, the tube morphology is created by the presence of a fluoride species and the formation and dissolution of a fluoride rich layer at the bottom and around the tubes. In terms of tube growth, the formation of a fluoride-rich layer at the tube bottoms and walls can be explained by the fact that $F^-$ ions compete with $O_2^-$ inward migration (Fig. 1b) and that the rate of inward migration of $F^-$ ions is twice that of $O^{2-}$.[130] As a result, a $TiF_4$ (or oxyfluoride) layer forms at the metal/oxide interface. As anodization proceeds, this layer also decorates the tube walls (Fig. 3c).[103,123] Experimental evidence for the formation of a distinct fluoride-rich layer at the bottom and in-between the $TiO_2$ nanotubes has been provided by means of high resolution scanning Auger Electron Spectroscopy (AES) and TEM analysis with elemental mapping. AES allows a detailed mapping of compositional variations (both horizontal and vertical mappings, Fig. 4a,b) across self-organized TiO$_2$ nanotube layers grown in fluoride containing organic-based electrolytes.[131,132] Concentration scans of fluoride signal are superimposed on the original SEM images for both horizontal (Fig. 4a) and vertical (Fig. 4b) modes. Clearly, the fluoride concentration peaks in-between individual tubes (Fig. 4a) and at the bottom of tubes (Fig. 4b). Moreover, Ti and O signals recorded in horizontal mapping mode follow the same trend as F, while C-rich regions correspond to remnants in the inner tube shell (Fig. 4a).[131] High-angle annular dark field scanning TEM (Fig. 4c) and elemental mapping (Fig. 4d) also highlight that the distribution of F element is higher approaching the tube walls, which confirms the comparably higher concentration of F$^-$ in-between tubes.[132] As Ti fluorides or oxy-fluorides are water soluble, this fluoride-rich layer is prone to chemical dissolution in aqueous electrolytes. As a result, this layer is permanently etched out – the faster, the higher the water content in the electrolyte, and this leads to a transition of nanopores-into-nanotubes. This pore-wall-splitting is apparent in Fig. 3c and in line with various experimental findings. In general (for various porous/tubular oxides), the solubility of the cell walls in an anodizing electrolyte is the critical factor that decides whether self-ordered oxide tubes or pores are formed.[31,113] Firstly, with a decreasing water content in an ethylene glycol electrolyte at a distinct concentration of 0.7 vol.% H$_2$O, the dissolution stops and a nanoporous TiO$_2$ layer is obtained rather than a tube layer.[113] Thus, water not only plays a crucial role during anodization being the source of oxygen to grow the oxide, but it is also an essential factor for the formation of tubes rather than pores. Accordingly, for Al anodized in acidic electrolytes a self-organized porous nanostructure is obtained.[30,31] But even if alumina is anodized in fluoride solutions, pores (not tubes) are observed as Al forms water-insoluble fluoride compounds.[31] Also the effect of an increased F$^-$ concentration in the electrolyte (0.025 M – 0.20 M) was found in line with the incorporation into the cell boundaries of TiF$_4$ species, as it leads to clearly larger splitting of adjacent pores.[113] These observations highlight that transitions from porous to tubular morphologies are gradual, with decisive factor being the solubility of cell boundaries in the anodization electrolyte. However, if anodization is carried out in water-based electrolytes, the dissolution of the fluoride is so high that splitting down the tubes can be faster than growth of tubes. This leads to a rippled morphology of the walls of those tubes (as we discuss below).

Noteworthy, by thermal annealing in air (typically performed at 450–500°C) not only amorphous-into-crystalline conversion of $TiO_2$ nanotubes is achieved, but also a loss of the fluoride-rich barrier layer at tube bottoms occurs. This may be ascribed to the relatively low sublimation temperature of $TiF_4$ (*i.e.*, 285°C).[133]

### 3. Various tube morphologies

Above we discussed mainly the formation of straight, smooth-walled tubes. Variations in experimental conditions can yield specific morphologies, such as the occurrence of side wall ripples,[51] bamboo-type tubes,[55] double/single-walled tubes,[53,54] top-open/partially top-closed tubes,[134–136] branched-tubes,[56] inter-tubes[57,137] and tubes-in-tubes.[132] The formation of ripples along tube walls is a result of the competition between tube growth and splitting speed.[51,138,139] When the fluoride rich layer between the tubes is dissolved faster than the tubes grow, this leads to a situation where the electrolyte between the tubes reaches the underneath metal. This in turn causes a local rapid oxide formation (a repassivation spike) and a sealing of this point by oxide – a rib is formed. For some time, this point seals the intertube space while the tubes continue to grow, until the oxide "seal" (the rib) gets penetrated by the etching action of the electrolyte and a rapid dissolution of the fluoride-rich layer between the tubes occurs again. The repetitive auto-sealing/dissolution cycle leads to periodic ribs (Fig. 5a) that can be accompanied by minor regular current oscillations recorded during anodization experiments. If the water content in the electrolyte is sufficiently reduced, such as for nanotubes grown in typical organic-based electrolyte (glycerol, ethylene glycol, DMSO), no detectable current fluctuations are observed and smooth tubes are obtained.[51] Other variations in morphology can be caused by alternation of applied voltage during anodization. This is based on the fact that the tube diameter of ordered nanotubular layers is closely related with the anodic growth factor ($f$) of the metal. Assuming that anodization starts from a frontier spot on the metal surface (*e.g.*, an oxide breakdown site), the oxide grows in all directions starting from this source point and a hemispherical shaped dome with radius $r$, given by $r = f \times \Delta U$, will form. This, to a large extent, determines the diameter of growing tubes. Diameter control is of great significance for various applications of $TiO_2$ tubes, *e.g.* when nanotube layers are used as morphology-directed template, hierarchical structure, size-selective flow-through membrane, or for their interaction with living matter in biological applications.[31,105,109,140–142]

Bamboo-type tubes,[138,139] branched and inter-tubes,[56,57,143] as well as tube stacks[80,82,137,144] can be grown by a stop and continuation of the anodization, or pulsing the voltage during anodizing. By using appropriate alternating voltage cycling, a morphology that resembles bamboo is formed (Fig. 5b). With this approach, oxide that connects tubes at their bottom is grown at each voltage step and by using different anodization durations the distance between the stratification layers can be adjusted.[55] By reducing the voltage to distinct lower voltage,[56,57] also branched tubes can be grown with distinct lower tube diameters. Depending on the anodization conditions, the second tube layer is either grown through the bottom of the first layer (Fig. 5c)[56] or in the gaps between the existing tubes of the upper layer, that is, inter-tubes are formed (Fig. 5d).[57,143] Heteromaterial superlattices can be produced using sputter-deposited multilayer metal stacks as substrate for anodization – this has been shown for alternating metal layers (*e.g.*, Ti–Ta, Ti–Nb, Ti–Zr).[80,82,144] Key is that experimental conditions can be established that lead for both elements to a self-organized tubular nanostructure (Fig. 5e). Given the large number of materials for which self-organized oxide nanostructures can be grown, this technique can produce a wide palette of multilayered nanostructures. A peculiar morphology are tube-in-tube structures as shown in Fig. 5f,g that can be regarded as a special case of double-walled nanotubes as discussed below. More detailed descriptions of different tube morphologies can also be found in Refs. 51,56,100,145.

### 4. Double- vs. Single-walled

An important aspect of tube growth that is widely neglected or often overlooked in literature is the self-induced double-layered, or double-walled, morphology of anodic oxide nanotubes grown in many organic electrolytes, which for several applications is detrimental. We will discuss, in the following, the origin of these double-walled tubes (Fig. 6a–d) and ways to produce single-walled tubes (Fig. 6e,f) of a much higher oxide quality.[52–54,142,146]

The most widely used electrolytes to grow ordered and seemingly well-defined $TiO_2$ nanotubes are based on ethylene glycol, water and fluorides. While these tubes in top or side view SEM appear uniform, a close look at ion milled cross sections in SEM (Fig. 6a,b) or tubes in TEM (particularly at the lower tube end), reveals that the walls of these $TiO_2$ nanotubes consist of two different types of oxide, namely an outer shell (OS) and an inner shell (IS), as also illustrated in Fig. 3c. The two-shell nature becomes even more apparent when the tubes are annealed to crystallize their tube walls. If the annealing temperature is sufficiently high and/or for a more rapid annealing than ~10°C/min, a clear separation between inner and outer shells can no longer be observed as IS and OS sinter together (maintaining the lower quality oxide characteristics).[52–54,147] As discussed in more detail in Section 7, this can be a critical factor in affecting not only the geometry but also the properties of $TiO_2$ nanotube layers. Detailed investigations showed that the as-grown outer shell consists of comparably pure $TiO_2$, while the inner layer is composed of carbon-rich titanium oxide or oxy-hydroxide.[52,54,142,148] To a large extent, this $TiO_xH_yC_z$ layer stems from precipitated Ti-ions, ejected from the oxide but not solvatized as $TiF_6^{2-}$, combined with carbon species embedded from the electrolyte during the anodization process. This inner layer typically dictates the inner tube morphology as it narrows the inner diameter of the tubes and limits the possibilities of a uniform and functional coating of the inner tube wall.[52–54] This inner layer is more sensitive to etching in the fluoride-containing electrolyte, compared with the outer shell $TiO_2$. This results in a V-shaped inner tube morphology (Fig. 3c), while OS preserves a nearly constant thickness along the whole tube length. For tubes grown for some time, a proper evaluation of double wall formation can therefore only be done close to the bottom of the tubes, where the thickness of IS is still maintained (Fig. 6c). The type of electrolyte used has a strong impact on whether or not, or to what extent, an inner contamination shell is formed. As mentioned, the most frequently adopted ethylene glycol (EG)/$H_2O$ mixtures that contain $NH_4F$ or HF lead to double-walled nanotubes, whereas dimethyl sulfoxide (DMSO) or EG/DMSO mixtures can be applied to grow single-walled tubes. This effect can be ascribed to different reactions of the electrode (anode) with the adjacent electrolyte (DMSO *vs.* EG) during anodization. High anodic voltages cause Schottky barrier breakdown in the oxide, which generates valence band holes that, similarly to photocatalysis, can react with the environment, either directly with electrolyte species or with $H_2O$, and form $OH^•$ radicals.[54,149] A range of carbonaceous oxidation products can be produced that adsorb on $TiO_2$ and are incorporated in the OS. In contrast to EG, DMSO is capable of capturing radicals nearly 10-time faster[150] and the reaction products are solvatized in the electrolyte while, in the case of EG, they are incorporated in the IS.[53,54] More recent reports introduced an even more widely useful technique for fabricating single-walled $TiO_2$ nanotubes. It was found that for essentially all double-layered $TiO_2$ nanotubes, the IS can be selectively etched leaving behind the OS, as an intact single wall, by exposing the samples to a piranha solution under optimized etching conditions, as shown in Fig. 6d–f.[142] This has been demonstrated recently in detail for tubes grown from EG electrolytes containing $NH_4F$, and lactic acid, LA (here, LA acts as an additive that allows for an unprecedented ultrafast growth of ordered $TiO_2$ nanotubes due to a field supporting effect).[146] Noteworthy, the IS-removal process can be carried out successfully over the entire nanotubular layer, for a wide range of tube lengths *i.e.*, from 2.5 μm up to 16 μm as reported in Refs. 142,148.

A more detailed comparison of the composition of double-walled and single-walled tubes produced by selective etching of IS has been obtained by means of Thermo-Gravimetric Analysis, coupled with Mass Spectrometry (TGA-MS) – characteristic decomposition patterns are shown in Fig. 7a,b. It is apparent that ~ 2–3 wt.% loss occurs in both layers, owing to water desorption (notably, the desorption of physically adsorbed water occurs at ~ 180–200°C, due to the presence of hydroxyl groups that interact with uncoordinated surface atoms).[151] In contrast, a clear difference in $CO_2$ release signals can be observed: $CO_2$ evolution in single-walled tubes produces a relatively low weight loss (~ 5 wt.%) that can be reasonably explained with the pyrolysis of environmental contaminants that are adsorbed on $TiO_2$ surface,[53] while remarkably higher amount of carbon is released as $CO_{2(g)}$ if double-walled tubes are heated up to ~ 350–400°C. This is ascribed to a thermal oxidation of the carbon species in the inner shell of the tubes.[132]

Finally, the recently reported tube-in-tube morphology (Fig. 5f,g) is an unusual arrangement obtained under conditions similar to that for ultra-fast growth of double-walled $TiO_2$ nanotubes.[146] Though more detailed investigations are required, the formation of this unconventional morphology is ascribed to the low temperature adopted during the anodization process that both limits the tube growth speed and enables a particularly slow fluoride-etching of C-rich layers.[132]

## 5. Towards maximized ordering and ideal arrangements

Systematic experimental investigations of growth parameters led over the years to a remarkably improved control over the morphology of the tubes, as well as of the degree of self-organization. While the historically first used water-based electrolytes yield nanotube layers with a relatively poor degree of self-ordering,[46–49,51] today a variety of electrolytes and conditions[109,152,153] can provide virtually ideal hexagonal-order over long surface ranges – but typically ideal order is limited to the size of an individual Ti crystal grain.[154] For $TiO_2$ nanotubes, strategies that can provide an extended degree of self-organization have been adopted from approaches developed for ordered porous alumina, *e.g.*, the use of pure and large-grain substrates that reduce the detrimental effect of substrate contaminants, inclusions and grain boundaries on the degree of ordering.[155] A most facile fabrication technique for improved order is double-anodization. This multiple-anodization technique was introduced by Masuda *et al.* for anodic $Al_2O_3$.[40] The concept is based on the fact that the degree of self-organization in anodic layers is improved as the tubes grow, and gradually increases towards their bottom (misfitting tubes are eliminated). Therefore, if a first grown sacrificial nanotube layer is removed from the metallic substrate, hexagonally ordered footprints (dimples) are left behind that serve as pretextured initiation sites for the following anodization (Fig. 8a). The so-fabricated tubes feature a more regular hexagonal shape (thus, the standard deviation in the tube diameter is narrowed), and the number of defects is strongly decreased (Fig. 8b,c).[152,156–158]

However, long-range ordering, with a narrowly dispersed pore diameter distribution and a domain size larger than substrate grains, is fundamental for some emerging applications such as photonic crystals and high density magnetic storage media.[43,159,160] Strategies that can induce long-range surface pretexturing are mold imprinting or focused-ion-beam (FIB) surface nanopatterning. FIB-induced surface pretexturing[161] offers the possibility of producing ordered concave arrays as initiation sites for anodization, with uniform and controllable depth, that enables the fabrication of $TiO_2$ nanotubes with different patterns with square, triangular and flower-like cell shapes, also with accurate control on inter-tube distance (Fig. 8d).[162–164] Nevertheless, FIB patterning (pixel after pixel) is a time-consuming sequential writing process and thus inadequate for commercial patterning of large areas of a Ti metal substrate. Although, mask-based, and thus parallel ion implantation (ion milling) processes have been proposed that remain still widely unexplored.[165]

A straightforward parallel process to create initiation sites that keep the registry across Ti metal grains is nanoimprinting the Ti surface with an ordered array of convexes using a metal mold (Fig. 8e–g). In this case, the mold is produced using $e^-$ beam lithography. The advantage of this technique is that the mold can be used repeatedly, and spacing and patterns are adjustable by the lithography process that produces the mold. Most inhomogeneity typically observed under non-guided self-organizing anodization (such as initiation layers and the grain-dependent orientation dictated by the polycrystallinity of metallic substrates) is also largely eliminated. However, given the high Young's modulus of Ti (116 GPa), only molds of very hard materials can be adopted.[166,167]

## 6. Advantages of lifted-off $TiO_2$ nanotube layers

The defined geometry of nanotubes and the possibility of controlling their length and diameter make anodic tubes particularly suitable for membrane-type applications that additionally exploit $TiO_2$ functional features, such as flow-through microreactors for photocatalysis and dye-sensitized solar cells.[141,142,147,168–170] For solar cells, not only the ease of electrolyte access in a two side open membrane (Fig. 9a) is of advantage, but even more beneficial are the superior electronic properties (lower electrical resistance) provided by bottomless tubes.

Nanotube membranes are produced by separating the anodic tube layer from the metallic substrate, followed by tube bottom opening. Albu *et al.* reported the first self-organized, free-standing $TiO_2$ nanotube membrane prepared by selectively dissolving the Ti substrate with a metal etching treatment in a $CH_3OH/Br_2$ solution. After substrate dissolution, the tubes were still closed; thus exposing the layer to HF vapors was necessary in order to open the bottom.[141] Afterwards, a range of different approaches were reported for both lift-off and opening procedures, such as chemical etching, sonication, and voltage alteration.[171–173] The main challenge encountered when separating nanotubes from the substrate is due to the relative brittleness of the anodic layer, particularly if the membrane is thinner than 50 μm or when large membrane areas are desired. Also, bottom removal mainly by HF, oxalic acid, $H_2O_2$ or other etchants has to be carefully controlled.

Most reliable approaches have recently been developed that allow the detachment of $TiO_2$ nanotube layers from the Ti metal substrate to form self-standing membranes as thin as ~2 μm.[147,168,169,174–176] Here an anodization–annealing–$2^{nd}$anodization sequence is employed: following the first anodization step, the nanotube layers are annealed in air at a comparably low temperature, *i.e.*, 150–300°C (to make the tubes water-free and partially crystalline); then the second anodization is carried out to produce a thin underlayer of amorphous $TiO_2$ NTs. This second amorphous layer can then be selectively dissolved by a treatment in HF or $H_2O_2$ solutions. Due to the much lower etching rate of the annealed layer, the amorphous tubes are selectively dissolved and free-standing membranes, that are defect-free over the entire surface area, can be produced.[147,168,169,174,175] Noteworthy, only when the lift-off process is performed in $H_2O_2$, tube bottom opening is simultaneously achieved.[168] Alternatively, the first layer can be produced from a double anodized layer[152,156–158] that then is annealed and chemically undermined. This treatment has been applied to self-standing and stable $TiO_2$ NTs membranes of ~1.8 – 50 μm thickness.[176]

A unique possibility offered by bottomless $TiO_2$ NT membranes is the fabrication of size-selective flow-through microreactors that can be used for virtually any photocatalytic reaction (*e.g.*, pollutant degradation and membrane de-clogging in protein separation).[141,170,177] For example, if a tube membrane is used to separate two (differently filled) compartments of a flow-through reactor, a concentration gradient is created that governs the solution diffusion until an equilibrium is reached. Interestingly, if the surface of the membrane is illuminated, photocatalysis (and membrane de-clogging that is particularly useful for re-opening of the membrane channels)[177] can be activated that, by permanently degrading the arriving reactant molecules, contributes to accelerate the solution diffusion.[141,170]

Recently, also considerable benefits of using lift-off membranes as anodes for dye-sensitized solar cells (DSSCs) have been reported. To construct photo-anodes such membrane layers are, either with[147,169] or without[168] bottom, glued to a conducting glass surface using $TiO_2$ nanoparticle suspensions. These arrangements have several advantages: not only the photo-anodes provide the classic advantage of 1D-nanostructures, namely the one-dimensional transport, but additionally if the anodic $TiO_2$ NT layer is detached from the Ti substrate and transferred onto FTO, the photo-anode can be irradiated through the FTO glass in a "front-side" illumination configuration (Fig. 9a). With this geometry, the DSSC reaches higher efficiency as the light losses typically observed in a back-side application due to light absorption in the electrolyte and by the Pt counter electrode are minimized.[147,168,169,178] Moreover, the solar cell efficiency is significantly enhanced for bottom-opened tube membranes (Fig. 9b).[168] This has been ascribed to a higher electron mobility within the tube layer, owing to the absence of the high-resistivity tube bottom, and to an optimal electrolyte percolation through the dye-sensitized fully-open tube scaffold. This in turn minimizes charge recombination as, upon electron injection into $TiO_2$, the dye can readily react with the electrolyte in its vicinity; even more, when a bottomless tube configuration is adopted, the ability of the dye solution to penetrate the entire $TiO_2$ structure is significantly enhanced; this also implies that an open-bottom morphology allows for optimal dye-loading.

It should also be mentioned that the detachment of tubes from the metal prevents the formation of rutile at the metal/oxide interface during annealing (rutile is formed by thermal oxidation of Ti metal). Rutile is detrimental for $TiO_2$ in photo-electrochemical applications due to the poor electronic properties compared with anatase.[179,180] It is well established that direct annealing of metallic Ti in air at temperatures above 450°C leads to a relatively thick rutile layer – this also occurs with the metal underneath the tubes. If the metal layer is absent, as in detached membranes, no rutile is detected up to annealing temperature of 950°C (Fig. 9c,d).[169,170] Although no remarkable difference in the phase composition of $TiO_2$ membranes annealed up to such a high temperature can be observed, it is evident that a gradual increase in the crystallinity degree, that is, a more intense main anatase reflection (at $2\theta$ ~25.3°), is obtained when increasing the annealing temperature up to 650°C.[170] This in turn enhances the charge transport properties of the layer as electron transport in the tubes becomes considerably faster (this is remarkably advantageous for maximizing charge collection at the oxide/electrode interface in DSSC).[169] Finally, also annealing in $O_2$ produces fully anatase $TiO_2$ membranes up to 750°C (that is, no rutile is detected), if the nanotube layers are detached from the Ti metallic substrate. As discussed below, $O_2$-annealing, for instance, is particularly useful to fill native oxygen vacancies in $TiO_2$ and, hence, produces a "higher quality" oxide, with optimized conductivity.[181]

## 7. Tube morphology and optical and electronic properties

Most relevant applications of TiO$_2$ nanotube layers ensue from the beneficial combination of the material's intrinsic properties with the one-dimensional nanoscale geometry. A key point is that this geometry can in principle provide a directional (*i.e.*, faster) charge transport through the oxide. In any form of TiO$_2$, another factor that strongly affects the electronic properties is the crystal structure of the material. In its amorphous form (such as in as-grown tubes), TiO$_2$ has a mobility gap of 3.2–3.5 eV and a very low electron mobility. Upon crystallization to anatase or rutile (typically achieved under air annealing conditions) an indirect optical band gap of 3.2 eV and 3.0 eV is obtained, respectively.[182] Both polymorphs show a higher mobility than amorphous TiO$_2$ but, in particular, a crystallization into anatase is often reported to be crucial to obtain the highest electron mobility.[183,184] Optical and electronic properties of TiO$_2$ are largely influenced by oxygen vacancies and Ti$^{3+}$ states in the lattice, that is, electrons may be trapped in the Ti$^{3+}$ states or holes in the oxygen vacancies and easily injected in the nearby conduction and valence bands by thermal excitation. The conductivity of TiO$_2$ is dominated by Ti$^{3+}$ trap states and, in the polycrystalline material, also by grain size. Therefore, a variation in the annealing conditions of TiO$_2$ nanotubes (such as temperature, time, and atmosphere) can alter the layer resistance (and the electron mobility) over several orders of magnitude.[185]

As previously highlighted, annealing of TiO$_2$ on the metal substrate (Ti) strongly differs from annealing free-standing membranes. In both cases, upon increasing the annealing temperature (*i.e.*, 250–500°C) a significant decrease in the tube resistivity can be obtained, owing to amorphous layers being converted to crystalline anatase. However, for TiO$_2$ nanotubes on the Ti substrate, annealing as described above may cause rutile formation due to the thermal oxidation of the metal substrate. As a negative consequence (in view of photo-electrochemical applications), the overall resistivity of these tube layers rises again (Fig. 10a).[186] In contrast, free-standing membranes can be annealed up to ~1000°C while maintaining a pure anatase phase – rutile is not formed owing to the absence of the Ti metal substrate (Fig. 9c,d), and such layers show clearly improved electronic properties.[147,169,170,181]

Typical resistivity values in the range of $10^4$–$10^5$ $\Omega$ cm[186] are obtained (from both four- and two-point measurements) for anatase tube layers anchored on the metal substrate– four-point measurements may however better reflect the resistivity dependence on the annealing temperature. However, two- and even four-point measurements carried out with metal contacts on the tubes, but the metal substrate underneath, are somewhat ambiguous. Most informative are four-point techniques applied to a single $TiO_2$ nanotube (Fig. 10b–d).[187] A crucial finding is that the conductivity measurements in this case show orders of magnitude lower values than measurements on the substrate. This large difference cannot be ascribed to the presence of the rutile underlayer but rather reflects an extremely high resistivity of the tube bottoms (this explains why bottom-opened tubes lead to significantly higher solar conversion efficiencies when used in DSSCs, Fig. 9b). Moreover, the T-dependence of resistance for a single tube was found compatible with a Mott variable range hopping mechanism, that is, conductivity is provided by electron hopping within localized states in a mobility gap.[187,188] In other words, these measurements indicate that not grain boundaries (crystal size) in the tube wall are limiting the conductivity, but the intrinsic defect states within $TiO_2$ crystals forming the nanotubes. While these measurements on single tubes are most informative, many data were obtained using the classic tubes-on-metal/two-point measurements – simply due to the facility of this approach. In spite of all the shortcomings of two-point measurements, phenomena such as water loss, anatase/rutile formation and defect-elimination by pure oxygen annealing can be detected and are in line with expectations.

Using this two-point approach, the effect of the inner C-rich layer on conductivity was also investigated, by comparing double-walled and single-walled NTs.[53] A clear difference in the conductivity can be already observed with as-formed tubes, with the double-layered exhibiting a somewhat lower resistance. However, after annealing, a drastic increase (about 10–100 times) in the conductivity properties is obtained for the single-walled tubes while a resistivity increase is observed for the double-walled tubes (Fig. 10e). Most likely the trend observed for double-walled tubes is due to the high amount of residual water that remains entrapped in the tubes up to ~180–200°C, as evident from TGA-MS results and, even more, can be ascribed to the high C-contamination content that originates from the IS layer (Fig. 7a);[53,54,132] the enhanced electrical conductivity of single-walled tubes therefore indicate that removal of the inner contamination layer leads to beneficial electrical properties in the tubes after crystallization (single-walled tubes show a pure (no contamination) and large crystalline $TiO_2$ structure, whereas in double-walled tubes contamination is still present and a typically smaller $TiO_2$ grain size within the tube wall is observed).[53,54,132] Nevertheless, even better electron transport properties can be achieved exposing single-walled tubes to a $TiCl_4$ treatment. Not only the tube walls become then decorated with a nanoparticle layer, but the treatment seems also able to passivate defect states, leading to clearly faster electron transport (Fig. 10f).[142,189]

However, the C-contaminated IS of double-walled tubes has also been used to create beneficial effects when annealing EG-based tubes under inert gas conditions. A thin carbon layer can be formed that uniformly coats the inner walls of $TiO_2$ NTs.[53,54,146,190] Such a cladding with a graphitic-like layer leads to enhanced conductivity, and provides a path to attach additional functionalities to the tubes, as the entire chemistry for modifying carbon-based layers becomes accessible[191,192] and applicable to $TiO_2$ NTs. This strategy can serve for example to fabricate scaffolds for anchoring a thin layer of $RuO_2$ nanoparticles (*i.e.*, a standard material for supercapacitors, Fig. 10g), and considerable performances of $TiO_2$ NTs as supercapacitor device can be obtained.[190]

Another promising way to improve electron transport in $TiO_2$, and hence its performance in photo-electrochemical applications (*e.g.*, DSSCs and photocatalysis), and to extend its light absorption spectrum in view of enabling visible-light photoactivity, is to introduce a dopant into the $TiO_2$ lattice. The dopant can either be a non-metal or a transition metal cation. In the first case, most frequently investigated are N,[193] C,[194] S,[195] and P[196] that either replace O-atoms in the $TiO_2$ structure (substitutional doping) or are stacked in interstitial spaces (interstitial doping), and generate suitable states in the band-gap and increase the conductivity of $TiO_2$ nanotubes (low dopant concentration) or lead to band-gap narrowing (high dopant concentration). Different approaches are reported to prepare doped $TiO_2$ nanostructures, among which thermal treatment in a suitable gas atmosphere, *e.g.* $NH_3$, is mostly adopted. Although findings of C-doping in $TiO_2$[25–27,197,198] are to a certain extent ambiguous,[199] C-containing tubes (that is, tubes with a double-walled morphology where carbon-uptake is due to the decomposition of the anodizing organic electrolyte under applied voltage) exhibit a change in the $TiO_2$ optical properties. Light absorption and photocurrent spectra taken for tubes and membranes show that carbon contaminants (~3 at.%) lead to a sub-band gap response (Fig. 10h). Also, the conversion of $TiO_2$ nanotubes to semimetallic $TiO_xC_y$, by thermal treatment in an acetylene atmosphere, or to TiN nanotubes, by a high temperature treatment in $NH_3$, drastically changes the electronic properties of the starting material to a semimetallic nature and also improves its mechanical stability.[200,201] Most recently, high pressure hydrogenation[202] as well as high-energy proton[203] and low-dose nitrogen[204] implantation into $TiO_2$ tube layers have been reported to alter not only the electronic structures of $TiO_2$, but also to create catalytically active surface sites for improving the photocatalytic $H_2$ evolution of $TiO_2$ NTs.

Finally, a most unique possibility offered by anodic $TiO_2$ nanotubes is the doping of tubes with a certain metal X, using suitable Ti-X alloys as metallic substrate for electrochemical anodization, where the alloyed species gets incorporated into the oxide. A number of reports show doping of tubes with Cr,[86] V,[77] W,[36] Ta[83–85] and Nb.[76,78,79,81,205] In particular, in optimally doped and annealed $TiO_2$ tubes, both Nb and Ta act as electron donors and hence increase the electron conductivity of the NT layers. This was *e.g.* demonstrated for nanotube arrays doped with small amounts (up to 0.1 at%) of Nb[78,79,205] or Ta,[84,85] for which a significantly enhanced performance in photo-electrochemical water splitting and dye-sensitized solar cells was observed. In contrast and interestingly, anodization of noble metal alloys of Ti lead to noble metal nanoparticle-decorated tubes,[89,90,206] *i.e.*, noble metals, such as Au and Pt, are not oxidized during the anodization process. With this approach, the decoration density of Au or Pt nanoparticles can be easily controlled by the amount of the noble metal in the alloy and by the anodization time. This tuning of nanoparticle interspacing is in turn particularly useful for obtaining an optimal (photo)catalytic activity, such as in the splitting of water or as chemo-selective catalytic sites.[89,90]

All these examples illustrate that for anodic $TiO_2$ NTs outstanding knowledge and control over a large set of inherent physico-chemical properties has been acquired within the last decade of research. These findings provide key strategies to tailor the characteristics of $TiO_2$ NTs, *e.g.*, their morphology, light absorption, conductivity, electron-hole mobility, *etc*. in a synergistic manner, in order to provide versatile platforms for various remarkable functional applications, such as photocatalytic and photo-electrochemical water splitting ($H_2$ generation), DSSCs, and size-selective microreactors for photocatalytic and biochemical reactions.[26,27,89,141,142,177,203,204]

**Conclusion**

Titanium dioxide is by far among the most investigated functional metal oxide. This is due to the set of specific properties (*e.g.*, non-toxicity, corrosion resistance, relatively long electron diffusion length, and biocompatibility) that makes $TiO_2$ attractive for various distinct applications including photocatalysis, photo-electrochemistry, and biochemistry. The possibility of growing porous anodic nanostructures and, in particular, self-organized and highly-aligned $TiO_2$ nanotubes through electrochemical anodization has further enlarged the research interest on this material due to other specific advantages, such as large surface area, directional charge and ion transport properties, size exclusion effects, and defined diffusion behavior.

The present review emphasizes critical aspects in the formation mechanism of self-ordered $TiO_2$ nanotubes, and wide range of morphology and property variations. We discuss the current understanding of the key factors that drive self-organization and attempt to show features common to ordered anodic nanoporous structures, such as porous $Al_2O_3$ and ordered $TiO_2$ nanotubes. In spite of common principles underlying their growth, there are crucial differences not only in terms of alumina and titania anodic nanostructure morphologies, but also the influence of various electrochemical growth parameters, that have been shown to strongly affect the morphology, composition and structure of the resulting nanotube layers. With this review we try to highlight the importance of such variations and, at the same time, to provide guidelines for the selection of growth conditions specific for the intended uses of the tube layers. We introduce most recent nanotube arrangements, such as stacks, bamboo-type tubes, branched and inter-tubes, tube-in-tube, but in particular we highlight key differences in single- and double-walled tubes, in their strongly distinct intrinsic properties. Moreover, we give an overview of the optimized preparation and use of $TiO_2$ tube membranes that in the recent few years have attracted strong interest in view of applications such as efficient DSSCs and size-selective flow-through photo-microreactors. Among most remarkable advantages of free-standing $TiO_2$ membranes is certainly the possibility of annealing the layers up to comparably high temperature (~1000°C) and obtaining the desired pure anatase phase (without formation of detrimental rutile). We also review the most relevant electronic and optical properties of differently grown $TiO_2$ nanotubes, giving a general overview on the approaches most frequently adopted to achieve the desired modifications. Owing to their intrinsic properties and to the versatility and scalability of the fabrication technique, $TiO_2$ nanotubes prepared through self-organizing anodization are nanostructured materials providing continuous surprises that trigger advances of nanotechnology and the development of energy-, environment- and bio-applications.

**Acknowledgments**

The authors would like to acknowledge the ERC, the DFG, and the DFG "Engineering of Advanced Materials" cluster of excellence for financial support. Prof. Dr. Radek Zbořil (Regional Centre of Advanced Technologies and Materials, Olomouc, Czech Republic), and Manuela Killian, Gihoon Cha, Marco Altomare, Anca Mazare, Jeong Eun Yoo, Xuemei Zhou and Ning Liu (Institute for Surface Science and Corrosion LKO, University of Erlangen-Nuremberg) are acknowledged for their valuable contributions.

# REFERENCES

1 C. M. Lieber, *Solid State Commun.*, 1998, **107**, 607–616.
2 C. Weisbuch and B. Vinter, *Quantum Semiconductor Structures: Fundamentals and Applications*, Amsterdam, 1991.
3 C. N. R. Rao, A. Muller and A. K. Cheetham, *The Chemistry of Nanomaterials: Synthesis, Properties and Applications.*, Weinheim, Germany, 2004.
4 S. Iijima, *Nature*, 1991, **354**, 56–58.
5 T. W. Ebbesen and P. M. Ajayan, *Nature*, 1992, **358**, 220–222.
6 S. Iijima and T. Ichihashi, *Nature*, 1993, **363**, 603–605.
7 D. T. Colbert, J. Zhang, S. M. McClure, P. Nikolaev, Z. Chen, J. H. Hafner, D. W. Owens, P. G. Kotula, C. B. Carter, J. H. Weaver, A. G. Rinzler and R. E. Smalley, *Science (80-. ).*, 1994, **266**, 1218–1222.
8 M. Bockrath, D. H. Cobden, P. L. McEuen, N. G. Chopra, A. Zettl, A. Thess and R. E. Smalley, *Science (80-. ).*, 1997, **275**, 1922–1925.
9 J. C. Charlier and P. Lambin, *Phys. Rev. B*, 1998, **57**, 15037–15039.
10 T. W. Odom, J.-L. Huang, P. Kim and C. M. Lieber, *Nature*, 1998, **391**, 62–64.
11 J. W. G. Wildöer, L. C. Venema, A. G. Rinzler, R. E. Smalley and C. Dekker, *Nature*, 1998, **391**, 59–62.
12 M. E. Spahr, P. Stoschitzki-Bitterli, R. Nesper, O. Haas and P. Novák, *J. Electrochem. Soc.*, 1999, **146**, 2780–2783.
13 F. Krumeich, H. J. Muhr, M. Niederberger, F. Bieri, B. Schnyder and R. Nesper, *J. Am. Chem. Soc.*, 1999, **121**, 8324–8331.
14 M. Remškar, Z. Škraba, P. Stadelmann and F. Lévy, *Adv. Mater.*, 2000, **12**, 814–818.
15 X. Sun and Y. Li, *Chem. - A Eur. J.*, 2003, **9**, 2229–2238.
16 D. V. Bavykin, J. M. Friedrich and F. C. Walsh, *Adv. Mater.*, 2006, **18**, 2807–2824.
17 B. D. Yao, Y. F. Chan, X. Y. Zhang, W. F. Zhang, Z. Y. Yang and N. Wang, *Appl. Phys. Lett.*, 2003, **82**, 281–283.
18 C.-J. Jia, L.-D. Sun, Z.-G. Yan, L.-P. You, F. Luo, X.-D. Han, Y.-C. Pang, Z. Zhang and C.-H. Yan, *Angew. Chem. Int. Ed.*, 2005, **44**, 4328–4333.
19 T. Kasuga, M. Hiramatsu, A. Hoson, T. Sekino and K. Niihara, *Langmuir*, 1998, **14**, 3160–3163.
20 W. Wang, O. K. Varghese, M. Paulose, C. A. Grimes, Q. Wang and E. C. Dickey, *J. Mater. Res.*, 2004, **19**, 417–422.
21 Y. Mao and S. S. Wong, *J. Am. Chem. Soc.*, 2006, **128**, 8217–8226.
22 P. Hoyer, *Langmuir*, 1996, **12**, 1411–1413.
23 M. S. Sander, M. J. Côté, W. Gu, B. M. Kile and C. P. Tripp, *Adv. Mater.*, 2004, **16**, 2052–2057.
24 M. S. Sander and H. Gao, *J. Am. Chem. Soc.*, 2005, **127**, 12158–12159.
25 G. Wu, T. Nishikawa, B. Ohtani and A. Chen, *Chem. Mater.*, 2007, **19**, 4530–4537.
26 P. Roy, S. Berger and P. Schmuki, *Angew. Chemie - Int. Ed.*, 2011, 50, 2904–2939.
27 K. Lee, A. Mazare and P. Schmuki, *Chem. Rev.*, 2014, **114**, 9385–6454.
28 A. Ghicov, S. Aldabergenova, H. Tsuchyia and P. Schmuki, *Angew. Chemie - Int. Ed.*, 2006, **45**, 6993–6996.
29 K. Yasuda and P. Schmuki, *Electrochim. Acta*, 2007, **52**, 4053–4061.
30 H. Tsuchiya, S. Berger, J. M. Macak, A. Ghicov and P. Schmuki, *Electrochem. commun.*, 2007, **9**, 2397–2402.

31  S. Berger, H. Tsuchiya and P. Schmuki, *Chem. Mater.*, 2008, 3245–3247.
32  L. Liu, W. Lee, Z. Huang, R. Scholz and U. Gösele, *Nanotechnology*, 2008, **19**, 335604.
33  H. Jha, R. Hahn and P. Schmuki, *Electrochim. Acta*, 2010, **55**, 8883–8887.
34  K. Lee and P. Schmuki, *Electrochem. commun.*, 2011, **13**, 542–545.
35  R. Ion, A. B. Stoian, C. Dumitriu, S. Grigorescu, A. Mazare, A. Cimpean, I. Demetrescu and P. Schmuki, *Acta Biomater.*, 2015, **24**, 370–377.
36  Y. C. Nah, N. K. Shrestha, D. Kim and P. Schmuki, *J. Appl. Electrochem.*, 2013, **43**, 9–13.
37  T. Rummel, *Zeitschrift für Phys.*, 1936, **99**, 518–551.
38  W. Baumann, *Zeitschrift für Phys.*, 1936, **102**, 59–66.
39  G. E. Thompson and G. C. Wood, in *Treatise on Materials Science and Technology*, Academic Press, New York, 1983, pp. 205–329.
40  H. Masuda and K. Fukuda, *Science (80-. ).*, 1995, **268**, 1466–1468.
41  K. Uosaki, K. Okazaki and H. Kita, *Anal. Chem.*, 1990, **62**, 652–656.
42  C. R. Martin, *Science (80-. ).*, 1994, **266**, 1961–1966.
43  K. Nielsch, R. B. Wehrspohn, J. Barthel, J. Kirschner, U. Gösele, S. F. Fischer and H. Kronmüller, *Appl. Phys. Lett.*, 2001, **79**, 1360–1362.
44  M. Steinhart, *Science (80-. ).*, 2002, **296**, 1997–1997.
45  G. . E. Thompson, *Thin Solid Films*, 1997, **297**, 192–201.
46  M. Assefpour-Dezfuly, C. Vlachos and E. H. Andrews, *J. Mater. Sci.*, 1984, **19**, 3626–3639.
47  V. Zwilling, E. Darque-Ceretti, A. Boutry-Forveille, D. David, M. Y. Perrin and M. Aucouturier, *Suface Interface Anal.*, 1999, **27**, 629–637.
48  D. Gong, C. a. Grimes, O. K. Varghese, W. Hu, R. S. Singh, Z. Chen and E. C. Dickey, *J. Mater. Res.*, 2001, **16**, 3331–3334.
49  R. Beranek, H. Hildebrand and P. Schmuki, *Electrochem. Solid-State Lett.*, 2003, **6**, B12–B14.
50  J. M. Macak, H. Tsuchiya and P. Schmuki, *Angew. Chemie - Int. Ed.*, 2005, **44**, 2100–2102.
51  J. M. Macak, H. Tsuchiya, L. Taveira, S. Aldabergerova and P. Schmuki, *Angew. Chemie - Int. Ed.*, 2005, **44**, 7463–7465.
52  S. P. Albu, A. Ghicov, S. Aldabergenova, P. Drechsel, D. LeClere, G. E. Thompson, J. M. Macak and P. Schmuki, *Adv. Mater.*, 2008, **20**, 4135–4139.
53  H. Mirabolghasemi, N. Liu, K. Lee and P. Schmuki, *Chem. Commun.*, 2013, **49**, 2067–9.
54  N. Liu, H. Mirabolghasemi, K. Lee, S. P. Albu, A. Tighineanu, M. Altomare and P. Schmuki, *Faraday Discuss.*, 2013, **164**, 107.
55  D. Kim, A. Ghicov, S. P. Albu and P. Schmuki, *J. Am. Chem. Soc.*, 2008, 16454–16455.
56  S. P. Albu, D. Kim and P. Schmuki, *Angew. Chemie - Int. Ed.*, 2008, **47**, 1916–1919.
57  X. Wang, L. Sun, S. Zhang and D. Zhao, *Electrochim. Acta*, 2013, **107**, 200–208.
58  S. P. Albu, A. Ghicov and P. Schmuki, *Phys. Status Solidi - Rapid Res. Lett.*, 2009, **3**, 64–66.
59  S. K. Mohapatra, S. E. John, S. Banerjee and M. Misra, *Chem. Mater.*, 2009, **21**, 3048–3055.
60  C. Y. Lee, L. Wang, Y. Kado, M. S. Killian and P. Schmuki, *ChemSusChem*, 2014, **7**, 934–940.
61  H. Tsuchiya, J. M. Macak, I. Sieber, L. Taveira, A. Ghicov, K. Sirotna and P. Schmuki, *Electrochem. commun.*, 2005, **7**, 295–298.
62  N. R. De Tacconi, C. R. Chenthamarakshan, G. Yogeeswaran, A. Watcharenwong, R. S. De Zoysa, N. a. Basit and K. Rajeshwar, *J. Phys. Chem. B*, 2006, **110**, 25347–25355.

63  I. Sieber, H. Hildebrand, A. Friedrich and P. Schmuki, *Electrochem. commun.*, 2005, **7**, 97–100.
64  W. Wei, K. Lee, S. Shaw and P. Schmuki, *Chem. Commun. (Camb).*, 2012, **48**, 4244–6.
65  Y. Yang, S. P. Albu, D. Kim and P. Schmuki, *Angew. Chemie - Int. Ed.*, 2011, **50**, 9071–9075.
66  C. Y. Lee, K. Lee and P. Schmuki, *Angew. Chemie - Int. Ed.*, 2013, **52**, 2077–2081.
67  I. V. Sieber and P. Schmuki, *J. Electrochem. Soc.*, 2005, **152**, C639.
68  W. Wei, J. M. Macak and P. Schmuki, *Electrochem. commun.*, 2008, **10**, 428–432.
69  W. Wei, J. M. Macak, N. K. Shrestha and P. Schmuki, *J. Electrochem. Soc.*, 2009, **156**, K104.
70  H. a. El-Sayed and V. I. Birss, *Nano Lett.*, 2009, **9**, 1350–1355.
71  W. J. Lee and W. H. Smyrl, *Electrochem. Solid-State Lett.*, 2005, **8**, B7.
72  S. Berger, F. Jakubka and P. Schmuki, *Electrochem. commun.*, 2008, **10**, 1916–1919.
73  S. Berger, J. Faltenbacher, S. Bauer and P. Schmuki, *Phys. Status Solidi - Rapid Res. Lett.*, 2008, **2**, 102–104.
74  H. Tsuchiya and P. Schmuki, *Electrochem. commun.*, 2005, **7**, 49–52.
75  S. Berger, F. Jakubka and P. Schmuki, *Electrochem. Solid-State Lett.*, 2009, **12**, K45–K48.
76  S. B. Aldabergenova, A. Ghicov, S. Albu, J. M. Macak and P. Schmuki, *J. Non. Cryst. Solids*, 2008, **354**, 2190–2194.
77  Y. Yang, D. Kim, M. Yang and P. Schmuki, *Chem. Commun.*, 2011, **47**, 7746–7748.
78  M. Yang, H. Jha, N. Liu and P. Schmuki, *J. Mater. Chem.*, 2011, **21**, 15205–15208.
79  C. Das, P. Roy, M. Yang, H. Jha and P. Schmuki, *Nanoscale*, 2011, **3**, 3094–3096.
80  M. Yang, G. Yang, E. Spiecker, K. Lee and P. Schmuki, *Chem. Commun.*, 2013, **49**, 460–462.
81  S. Ozkan, A. Mazare and P. Schmuki, *Electrochim. Acta*, 2015, **176**, 819–826.
82  H. Tsuchiya, T. Akaki, J. Nakata, D. Terada, N. Tsuji, Y. Koizumi, Y. Minamino, P. Schmuki and S. Fujimoto, *Electrochim. Acta*, 2009, **54**, 5155–5162.
83  W. Wei, S. Berger, N. Shrestha and P. Schmuki, *J. Electrochem. Soc.*, 2010, **157**, C409.
84  K. Lee and P. Schmuki, *Electrochem. commun.*, 2012, **25**, 11–14.
85  M. Altomare, K. Lee, M. S. Killian, E. Selli and P. Schmuki, *Chem. - A Eur. J.*, 2013, **19**, 5841–5844.
86  T. Mishra, L. Wang, R. Hahn and P. Schmuki, *Electrochim. Acta*, 2014, **132**, 410–415.
87  P. Agarwal, I. Paramasivam, N. K. Shrestha and P. Schmuki, *Chem. - An Asian J.*, 2010, **5**, 66–69.
88  P. Roy, C. Das, K. Lee, R. Hahn, T. Ruff, M. Moll and P. Schmuki, *J. Am. Chem. Soc.*, 2011, **133**, 5629–5631.
89  K. Lee, R. Hahn, M. Altomare, E. Selli and P. Schmuki, *Adv. Mater.*, 2013, **25**, 6133–6137.
90  S. N. Basahel, K. Lee, R. Hahn, P. Schmuki, S. M. Bawaked and S. a Al-Thabaiti, *Chem. Commun.*, 2014, **50**, 6123–5.
91  C. Y. Lee, L. Wang, Y. Kado, R. Kirchgeorg and P. Schmuki, *Electrochem. commun.*, 2013, **34**, 308–311.
92  L. Wang, C.-Y. Lee, R. Kirchgeorg, N. Liu, K. Lee, Š. Kment, Z. Hubička, J. Krýsa, J. Olejníček, M. Čada, R. Zbořil and P. Schmuki, *Res. Chem. Intermed.*, 2015, **41**, 9333–9341.
93  W. Li, J. Li, X. Wang, S. Luo, J. Xiao and Q. Chen, *Electrochim. Acta*, 2010, **56**, 620–625.
94  W. Wei, S. Shaw, K. Lee and P. Schmuki, *Chem. - A Eur. J.*, 2012, **18**, 14622–14626.

95  M. Altomare, O. Pfoch, A. Tighineanu, R. Kirchgeorg, K. Lee, E. Selli and P. Schmuki, *J. Am. Chem. Soc.*, 2015, **137**, 5646–5649.
96  A. Ghicov, H. Tsuchiya, J. M. MacAk and P. Schmuki, *Electrochem. commun.*, 2005, **7**, 505–509.
97  B. O'Regan and M. Grätzel, *Nature*, 1991, **353**, 737–740.
98  A. Hagfeldt, G. Boschloo, L. Sun, L. Kloo and H. Pettersson, *Chem. Rev.*, 2010, **110**, 6595–6663.
99  R. P. Lynch, A. Ghicov and P. Schmuki, *J. Electrochem. Soc.*, 2010, **157**, G76.
100 A. Ghicov and P. Schmuki, *Chem. Commun. (Camb).*, 2009, 2791–2808.
101 D. Kowalski, D. Kim and P. Schmuki, *Nano Today*, 2013, **8**, 235–264.
102 D. Regonini, C. R. Bowen, a. Jaroenworaluck and R. Stevens, *Mater. Sci. Eng. R Reports*, 2013, **74**, 377–406.
103 D. J. LeClere, a. Velota, P. Skeldon, G. E. Thompson, S. Berger, J. Kunze, P. Schmuki, H. Habazaki and S. Nagata, *J. Electrochem. Soc.*, 2008, **155**, C487.
104 A. Güntherschulze and H. Betz, *Zeitschrift für Phys.*, 1934, **92**, 367–374.
105 K. Yasuda, J. M. Macak, S. Berger, A. Ghicov and P. Schmuki, *J. Electrochem. Soc.*, 2007, **154**, C472–C478.
106 J. W. Schultze, M. M. Lohrengel and D. Ross, *Electrochim. Acta*, 1983, **28**, 973–984.
107 P. Schmuki, *J. Solid State Electrochem.*, 2002, **6**, 145–164.
108 L. V. Taveira, J. M. Macák, H. Tsuchiya, L. F. P. Dick and P. Schmuki, *J. Electrochem. Soc.*, 2005, **152**, B405.
109 J. M. Macak, H. Hildebrand, U. Marten-Jahns and P. Schmuki, *J. Electroanal. Chem.*, 2008, **621**, 254–266.
110 K. R. Hebert, S. P. Albu, I. Paramasivam and P. Schmuki, *Nat. Mater.*, 2011, **11**, 162–166.
111 Q. Van Overmeere, J.-F. Vanhumbeeck and J. Proost, *J. Electrochem. Soc.*, 2010, **157**, C166.
112 J. F. Vanhumbeeck and J. Proost, *Electrochim. Acta*, 2008, **53**, 6165–6172.
113 W. Wei, S. Berger, C. Hauser, K. Meyer, M. Yang and P. Schmuki, *Electrochem. commun.*, 2010, **12**, 1184–1186.
114 G. K. Singh, a. a. Golovin and I. S. Aranson, *Phys. Rev. B - Condens. Matter Mater. Phys.*, 2006, **73**, 1–12.
115 L. G. Stanton and a. a. Golovin, *Phys. Rev. B - Condens. Matter Mater. Phys.*, 2009, **79**, 1–7.
116 S. P. Albu, N. Taccardi, I. Paramasivam, K. R. Hebert and P. Schmuki, *J. Electrochem. Soc.*, 2012, **159**, H697–H703.
117 X. Zhou, N. T. Nguyen, S. Özkan and P. Schmuki, *Electrochem. commun.*, 2014, **46**, 157–162.
118 T. P. Hoar and N. F. Mott, *J. Phys. Chem. Solids*, 1959, **9**, 97–99.
119 V. P. Parkhutik and V. I. Shershulsky, *J. Phys. D. Appl. Phys.*, 2000, **25**, 1258–1263.
120 K. S. Raja, M. Misra and K. Paramguru, *Electrochim. Acta*, 2005, **51**, 154–165.
121 A. Valota, D. J. Leclere, T. Hashimoto, P. Skeldon, G. E. Thompson, S. Berger, J. Kunze and P. Schmuki, *Nanotechnology*, 2008, **19**, 355701.
122 S. Berger, J. Kunze, P. Schmuki, D. LeClere, A. T. Valota, P. Skeldon and G. E. Thompson, *Electrochim. Acta*, 2009, **54**, 5942–5948.
123 S. J. Garcia-Vergara, P. Skeldon, G. E. Thompson and H. Habazaki, *Electrochim. Acta*, 2006, **52**, 681–687.
124 S. P. Albu, A. Ghicov, J. M. Macak and P. Schmuki, *Phys. status solidi – Rapid Res. Lett.*, 2007, **1**, R65–R67.

125  O. Jessensky, F. Müller and U. Gösele, *Appl. Phys. Lett.*, 1998, **72**, 1173–1175.
126  P. Skeldon, G. E. Thompson, S. J. Garcia-Vergara, L. Iglesias-Rubianes and C. E. Blanco-Pinzon, *Electrochem. Solid-State Lett.*, 2006, **9**, B47.
127  K. Lee, J. Kim, H. Kim, Y. Lee, Y. Tak, D. Kim and P. Schmuki, *J. Korean Phys. Soc.*, 2009, **54**, 1027.
128  S. P. Albu and P. Schmuki, *Electrochim. Acta*, 2013, **91**, 90–95.
129  S. Berger, J. M. Macak, J. Kunze and P. Schmuki, *Electrochem. Solid-State Lett.*, 2008, **11**, C37.
130  H. Habazaki, K. Fushimi, K. Shimizu, P. Skeldon and G. E. Thompson, *Electrochem. commun.*, 2007, **9**, 1222–1227.
131  S. Berger, S. P. Albu, F. Schmidt-Stein, H. Hildebrand, P. Schmuki, J. S. Hammond, D. F. Paul and S. Reichlmaier, *Surf. Sci.*, 2011, **605**, L57–L60.
132  S. So and P. Schmuki, *to be Submitt.*, 2016.
133  I. Barin, *Thermochemical data of pure substances*, VCH, Weinheim, Germany, 3rd edn., 1995.
134  C. J. Lin, W. Y. Yu and S. H. Chien, *J. Mater. Chem.*, 2010, **20**, 1073–1077.
135  J. Choi, S.-H. Park, Y. S. Kwon, J. Lim, I. Y. Song and T. Park, *Chem. Commun.*, 2012, **48**, 8748–50.
136  J. Choi, Y. Y. S. Kwon and T. Park, *J. Mater. Chem. A*, 2014, **2**, 14380.
137  K. Yasuda and P. Schmuki, *Electrochem. commun.*, 2007, **9**, 615–619.
138  a. Valota, D. J. LeClere, P. Skeldon, M. Curioni, T. Hashimoto, S. Berger, J. Kunze, P. Schmuki and G. E. Thompson, *Electrochim. Acta*, 2009, **54**, 4321–4327.
139  Y. Y. Song and P. Schmuki, *Electrochem. commun.*, 2010, **12**, 579–582.
140  S. Bauer, S. Kleber and P. Schmuki, *Electrochem. commun.*, 2006, **8**, 1321–1325.
141  S. P. Albu, A. Ghicov, J. M. Macak, R. Hahn and P. Schmuki, *Nano Lett.*, 2007, **7**, 1286–1289.
142  S. So, I. Hwang and P. Schmuki, *Energy Environ. Sci.*, 2015, **8**, 849–854.
143  Z. Su and W. Zhou, *Sci. Found. China*, 2008, **16**, 36–53.
144  W. Wei, H. Jha, G. Yang, R. Hahn, I. Paramasivam, S. Berger, E. Spiecker and P. Schmuki, *Adv. Mater.*, 2010, **22**, 4770–4774.
145  S. Berger, R. Hahn, P. Roy and P. Schmuki, *Phys. Status Solidi Basic Res.*, 2010, **247**, 2424–2435.
146  S. So, K. Lee and P. Schmuki, *J. Am. Chem. Soc.*, 2012, **134**, 11316–11318.
147  F. Mohammadpour, M. Moradi, K. Lee, G. Cha, S. So, A. Kahnt, D. M. Guldi, M. Altomare and P. Schmuki, *Chem. Commun.*, 2015, **51**, 1631–1634.
148  I. Hwang, S. So, M. Mokhtar, A. Alshehri, S. a. Al-Thabaiti, A. Mazare and P. Schmuki, *Chem. - A Eur. J.*, 2015, **21**, 9204–9208.
149  Y. Y. Song, P. Roy, I. Paramasivam and P. Schmuki, *Angew. Chemie - Int. Ed.*, 2010, **49**, 351–354.
150  W. Rosenblum, *Ann. N. Y. Acad. Sci.*, 1983, **411**, 110–119.
151  U. Thanganathan and M. Nogami, *Materials Challenges in Alternative and Renewable Energy*, Wiley- VCH, 2010.
152  J. M. Macak, S. Albu, D. H. Kim, I. Paramasivam, S. Aldabergerova and P. Schmuki, *Electrochem. Solid-State Lett.*, 2007, **10**, K28.
153  S. P. Albu and P. Schmuki, *Phys. Status Solidi - Rapid Res. Lett.*, 2010, **4**, 215–217.
154  J. E. Yoo, K. Lee, M. Altomare, E. Selli and P. Schmuki, *Angew. Chemie - Int. Ed.*, 2013, **52**, 7514–7517.

155 C. Lee and P. Schmuki, in *Electrochemical Engineering Across Scales: From Molecules to Processes*, eds. R. C. Alkire, P. N. Bartlett and J. Lipkowski, Wiley-VCH, First Edit., 2015, pp. 145–192.
156 J. M. Macak, S. P. Albu and P. Schmuki, *Phys. Status Solidi - Rapid Res. Lett.*, 2007, **1**, 181–183.
157 G. Zhang, H. Huang, Y. Zhang, H. L. W. Chan and L. Zhou, *Electrochem. commun.*, 2007, **9**, 2854–2858.
158 A. Mazzarolo, K. Lee, A. Vicenzo and P. Schmuki, *Electrochem. commun.*, 2012, **22**, 162–165.
159 H. Masuda, M. Ohya, H. Asoh, M. Nakao, M. Nohtomi and T. Tamamura, *Jpn. J. Appl. Phys.*, 1999, **38**, L1403–L1405.
160 H. Masuda, M. Ohya, H. Asoh and K. Nishio, *Japanese J. Appl. Physics, Part 2 Lett.*, 2001, **40**, 2–5.
161 P. Schmuki and L. E. Erickson, *Phys. Rev. Lett.*, 2000, **85**, 2985–2988.
162 B. Chen and K. Lu, *Langmuir*, 2011, **27**, 12179–12185.
163 B. Chen, K. Lu and J. A. Geldmeier, *Chem. Commun. (Camb).*, 2011, **47**, 10085–7.
164 H. Amani Hamedani, S. W. Lee, A. Al-Sammarraie, Z. R. Hesabi, A. Bhatti, F. M. Alamgir, H. Garmestani and M. a. Khaleel, *ACS Appl. Mater. Interfaces*, 2013, **5**, 9026–9033.
165 A. Spiegel, W. H. Bruenger, C. Dzionk and P. Schmuki, *Microelectron. Eng.*, 2003, **67-68**, 175–181.
166 J. Choi, R. B. Wehrspohn, J. Lee and U. Gösele, *Electrochim. Acta*, 2004, **49**, 2645–2652.
167 T. Kondo, S. Nagao, T. Yanagishita, N. T. Nguyen, K. Lee, P. Schmuki and H. Masuda, *Electrochem. commun.*, 2015, **50**, 73–76.
168 F. Mohammadpour, M. Moradi, G. Cha, S. So, K. Lee, M. Altomare and P. Schmuki, *ChemElectroChem*, 2015, **2**, 204–207.
169 F. Mohammadpour, M. Altomare, S. So, K. Lee, M. Mokhtar, A. Alshehri, S. a Al-Thabaiti and P. Schmuki, *Semicond. Sci. Technol.*, 2016, **31**, 014010.
170 S. So and P. Schmuki, *to be Submitt.*, 2016.
171 C.-J. Lin, W.-Y. Yu, Y.-T. Lu and S.-H. Chien, *Chem. Commun. (Camb).*, 2008, 6031–6033.
172 K. Kant and D. Losic, *Phys. Status Solidi - Rapid Res. Lett.*, 2009, **3**, 139–141.
173 J. Wang and Z. Lin, *Chem. Mater.*, 2008, **20**, 1257–1261.
174 Q. Chen and D. Xu, *J. Phys. Chem. C*, 2009, **113**, 6310–6314.
175 M. Dubey, M. Shrestha, Y. Zhong, D. Galipeau and H. He, *Nanotechnology*, 2011, **22**, 285201.
176 G. Cha, P. Schmuki and M. Altomare, *Chem. - An Asian J.*, 2016, n/a–n/a.
177 P. Roy, T. Dey, K. Lee, D. Kim, B. Fabry and P. Schmuki, *J. Am. Chem. Soc.*, 2010, **132**, 7893–7895.
178 J. Choi, S. Song, G. Kang and T. Park, *ACS Appl. Mater. Interfaces*, 2014, **6**, 15388–15394.
179 A. Ghicov, S. P. Albu, R. Hahn, D. Kim, T. Stergiopoulos, J. Kunze, C. A. Schiller, P. Falaras and P. Schmuki, *Chem. - An Asian J.*, 2009, **4**, 520–525.
180 P. Roy, D. Kim, K. Lee, E. Spiecker and P. Schmuki, *Nanoscale*, 2010, **2**, 45–59.
181 S. So and P. Schmuki, *to be Submitt.*, 2016.
182 M. R. Hoffmann, S. T. Martin, W. Choi and D. W. Bahnemannt, *Chem. Rev.*, 1995, **95**, 69–96.

183 L. Forro, O. Chauvet, D. Emin, L. Zuppiroli, H. Berger and F. Lévy, *J. Appl. Phys.*, 1994, **75**, 633–635.
184 H. Tang, K. Prasad, R. Sanjines, P. E. Schmid and F. Levy, *J. Appl. Phys.*, 1994, **75**, 2042–2047.
185 S. So and P. Schmuki, *Angew. Chemie - Int. Ed.*, 2013, **52**, 7933–7935.
186 A. Tighineanu, T. Ruff, S. Albu, R. Hahn and P. Schmuki, *Chem. Phys. Lett.*, 2010, **494**, 260–263.
187 M. Stiller, J. Barzola-Quiquia, I. Lorite, P. Esquinazi, R. Kirchgeorg, S. P. Albu and P. Schmuki, *Appl. Phys. Lett.*, 2013, **103**, 173108.
188 N. F. Mott, *J. Non. Cryst. Solids*, 1968, **1**, 1–17.
189 K. Ishibashi, R. Yamaguchi, Y. Kimura and M. Niwano, *J. Electrochem. Soc.*, 2008, **155**, K10.
190 Z.-D. Gao, X. Zhu, Y.-H. Li, X. Zhou, Y.-Y. Song and P. Schmuki, *Chem. Commun.*, 2015, **51**, 7614–7617.
191 V. Singh, D. Joung, L. Zhai, S. Das, S. I. Khondaker and S. Seal, *Prog. Mater. Sci.*, 2011, **56**, 1178–1271.
192 T. Kuila, S. Bose, A. K. Mishra, P. Khanra, N. H. Kim and J. H. Lee, *Prog. Mater. Sci.*, 2012, 57, 1061–1105.
193 R. Asahi, T. Morikawa, T. Ohwaki, K. Aoki and Y. Taga, *Science (80-. ).*, 2001, **293**, 2000–2002.
194 T. Tachikawa, S. Tojo, K. Kawai, M. Endo, M. Fujitsuka, T. Ohno, K. Nishijima, Z. Miyamoto and T. Majima, *J. Phys. Chem. B*, 2004, **108**, 19299–19306.
195 T. Ohno, T. Mitsui and M. Matsumura, *Chem. Lett.*, 2003, **32**, 364–365.
196 K. Yang, Y. Dai and B. Huang, *J. Phys. Chem. C*, 2007, **111**, 18985–18994.
197 H. Wang and J. P. Lewis, *J. Phys. Condens. Matter*, 2006, **18**, 421–434.
198 J. H. Park, S. Kim and A. J. Bard, *Nano Lett.*, 2006, **6**, 24–28.
199 A. B. Murphy, *Sol. Energy Mater. Sol. Cells*, 2008, **92**, 363–367.
200 R. Hahn, F. Schmidt-Stein, J. Sahnen, S. Thiemann, Y. Song, J. Kunze, V. P. Lehto and P. Schmuki, *Angew. Chemie - Int. Ed.*, 2009, **48**, 7236–7239.
201 F. Schmidt-Stein, S. Thiemann, S. Berger, R. Hahn and P. Schmuki, *Acta Mater.*, 2010, **58**, 6317–6323.
202 N. Liu, C. Schneider, D. Freitag, U. Venkatesan, V. R. R. Marthala, M. Hartmann, B. Winter, E. Spiecker, A. Osvet, E. M. Zolnhofer, K. Meyer, T. Nakajima, X. Zhou and P. Schmuki, *Angew. Chemie Int. Ed.*, 2014, **53**, 14201–14205.
203 N. Liu, V. Häublein, X. Zhou, U. Venkatesan, M. Hartmann, M. Ma??Kovi??, T. Nakajima, E. Spiecker, A. Osvet, L. Frey and P. Schmuki, *Nano Lett.*, 2015, **15**, 6815–6820.
204 X. Zhou, V. Häublein, N. Liu, N. T. Nguyen, E. M. Zolnhofer, H. Tsuchiya, M. S. Killian, K. Meyer, L. Frey and P. Schmuki, *Angew. Chemie Int. Ed.*, 2016, n/a–n/a.
205 M. Yang, D. Kim, H. Jha, K. Lee, J. Paul and P. Schmuki, *Chem. Commun. (Camb).*, 2011, **47**, 2032–2034.
206 S. J. Garcia-Vergara, M. Curioni, F. Roeth, T. Hashimoto, P. Skeldon, G. E. Thompson and H. Habazaki, *J. Electrochem. Soc.*, 2008, **155**, C333.

**Captions**

**Figure 1.** a) Typical 2-electrode electrochemical anodization set-up and possible anodic morphologies: EP) metal electropolishing, CO) formation of a compact anodic oxide, PO) self-ordered nanostructured oxide (nanopores or nanotubes). b) High field oxide formation in the presence of fluoride ions: a steady-state is established between the oxide formation at the inner interface and its dissolution at the outer interface (due to dissolution/complexation of $Ti^{4+}$ as $TiF_6^{2-}$). Rapid fluoride migration leads to formation of a fluoride-rich layer at the $Ti/TiO_2$ interface. c) Typical current-time ($j$–$t$) characteristics and different stages in nanotube formation: (i) initial formation of a compact oxide layer; (ii) initiation of irregular nanoscale pores under the "right" anodizing conditions; (iii) formation of a regular nanoporous/nanotubular layer. d) Time evolution of stress×thickness ($\sigma_{ox}h_{ox}$, left axis) and cell voltage (right axis) during self-organizing anodization of a Ti thin film under galvanostatic condition. The inset shows the instantaneous roll-up of a thin NT layer (~ 200 nm) detached from the metallic substrate that is indicative of a residual compressive stress (scale bar of the SEM image 5 nm).

**Figure 2.** SEM images illustrating the structure of ordered oxide nanotube and nanopore layers electrochemically grown on different valve metals. Al: Reproduced with permission from Ref. 32. Ti: Reproduced from Ref. 142 with permission from the Royal Society of Chemistry; inset: Reproduced with permission from Ref. 146. W: Reproduced with permission from Ref. 95. Nb: Reproduced from Ref. 64 with permission from the Royal Society of Chemistry. V: Reproduced with permission from Ref. 65. Co: Reproduced with permission from Ref. 66. Ta: Reproduced with permission from Ref. 34.

**Figure 3.** a) TEM micrograph of an anodic alumina film after anodization, where "ridges" and "valleys" are highlighted. Reprinted with permission from J. Electrochem. Soc. **157** (11), C399 (2010). Copyright 2010, The Electrochemical Society. b) TEM image of $Al_2O_3$ nanopores with a tungsten tracer layer. Reprinted with permission from Electrochem. Solid-State Lett. **9** (11), B47 (2006). Copyright 2006, The Electrochemical Society. c) Schematic of $TiO_2$ nanotube formation: displacement of the fluoride-rich layer towards the cell boundaries by a plastic flow mechanism (left-side) and dissolution of the fluoride-rich cell boundaries and consequent formation of nanotubes (right-side). d) Schematic of the field concentration effects at the bottom of anodic nanotubes and model for length expansion: stress pushes some of the oxide up in the direction of tube walls. Reprinted with permission from J. Electrochem. Soc. **154** (9), C472 (2007). Copyright 2007, The Electrochemical Society.

**Figure 4.** High resolution Auger Electron Spectroscopy (AES) scans of different elements (Ti, O, C and F) recorded for a $TiO_2$ nanotube layer in (a) horizontal and (b) vertical modes. c) High angle annular dark field scanning TEM (HAADF-STEM) image and d) F elemental mapping of a double-walled $TiO_2$ nanotube. (a) Reproduced with permission from Ref. 131.

**Figure 5.** SEM images of: a) $TiO_2$ nanotubes with a rippled-wall morphology, b) bamboo-type tubes, c) branched-tubes, d) inter-tubes, e) $TiO_2$ and $Ta_2O_5$ superlattice tubular oxide. f), g) top and cross sectional views, respectively, of tubes-in-tubes. (a) Reproduced with permission from Ref. 138. (b,c) Reproduced with permission from Ref. 56. (d) Reproduced with permission from Ref. 57.

**Figure 6.** Cross-sectional SEM images of ion-milled (a) "as formed" and (b) annealed tubes (showing the typical double-walled morphology). c) Full cross-sectional SEM image of a double-walled $TiO_2$ nanotube layer (formed in EG-based electrolyte) with top-view SEM images taken at the top, from the fractures in the middle, and close to the bottom of a tube layer. d) Schematic drawing of inner layer removal through chemical etching with a 1:3 solution mixture of $H_2O_2/H_2SO_4$ for 6 min at 70°C. e) and f) Single-walled tubes after core removal process leaving behind only the outer tube shell. (a,b,d–f) Reproduced from Ref. 142 with permission from the Royal Society of Chemistry. (c) Reproduced from Ref. 54 with permission from the Royal Society of Chemistry.

**Figure 7.** Thermal desorption profile for $m/e^- = 18$ ($H_2O$) and for $m/e^- = 44$ ($CO_2$) measured by Thermo-Gravimetric Analysis coupled with Mass Spectrometry (TGA-MS) for (a) double-walled and (b) single-walled tubes.

**Figure 8.** a) Schematic of the fabrication of $TiO_2$ nanotubes by a double-anodization approach. SEM bottom views of ordered domains of $TiO_2$ NTs after the 1st (b) and 2nd (c) anodization. d) SEM image of $TiO_2$ nanotubes grown by FIB-guided anodization on a concave pattern. e) Schematic diagram of the fabrication of an ideally ordered anodic $TiO_2$ layer on a Ti foil pretextured through nanoindentation. f), g) Typical SEM micrographs of long-range ordered anodic titania nanotubes grown on a nanoindented Ti metal substrate. (b,c) Reproduced with permission from Ref. [156]. (d) Reproduced with permission from Ref. [162]. (e–g) Reproduced with permission from Ref. [167].

**Figure 9.** a) Sketch of anodic TiO$_2$ nanotube-based front-side illuminated DSSC. TiO$_2$ tube arrays are detached from Ti substrates, bottom-opened, and transferred onto FTO slides in a tube-top-down configuration. b) Dye loading, short-circuit current, and power-conversion efficiency of DSSCs fabricated from tube membranes transferred onto FTO glass in a tube-top-down configuration. The SEM images show a close view of the different degrees of tube bottom opening. Reproduced with permission from Ref. [168]. c), d) XRD patterns of membranes annealed at different temperatures and under different atmospheres: (c) 450–1050°C in air; (d) 450–750°C in O$_2$ (the patterns were collected after transferring the annealed membranes on quartz glass slides).

**Figure 10.** a) Comparison of specific resistivity values of TiO$_2$ nanotubes *vs.* annealing temperature obtained by 4-point (left axis) and 2-point (right axis) measurements. Reproduced with permission from Ref. [186]. (b–d) Procedure of four-point conductivity measurement with a single nanotube: (b) Single TiO$_2$ nanotube before fixing it on the substrate, (c) the nanotube fixed with WC$_x$, and (d) after producing the electrical contacts. Reproduced with permission from Ref. [187]. e) Comparison of electrical resistance of the single-walled and double-walled nanotube layers measured at different temperatures. Reproduced from Ref. 53 with permission from the Royal Society of Chemistry. f) Electron transfer time ($\tau_c$) constants obtained from IMPS measurements performed under UV light irradiation (325 nm) for double-walled (Double) and single-walled (Single) TiO$_2$ nanotubes, also modified with a layer of TiCl4 nanoparticles (Double+T1 and Single+T1, respectively). Reproduced from Ref. 142 with permission from the Royal Society of Chemistry. g) TEM micrographs of bare TiO$_2$ NTs before annealing (left-side) and of carbon-cladded TiO$_2$ NTs modified with the deposition of RuO$_2$ nanoparticles (right-side). Reproduced from Ref. [190] with permission from the Royal Society of Chemistry. h) IPCE spectra of different

C-contaminated TiO$_2$ nanotube layers that show sub-band gap response: (■) amorphous NTs, (●) NTs annealed in air, (○) NTs annealed in CH$_3$OH-containing atmosphere. Reprinted with permission from J. Electrochem. Soc. **157** (3), G76 (2010). Copyright 2010, The Electrochemical Society.

**Figures**

Figure 1

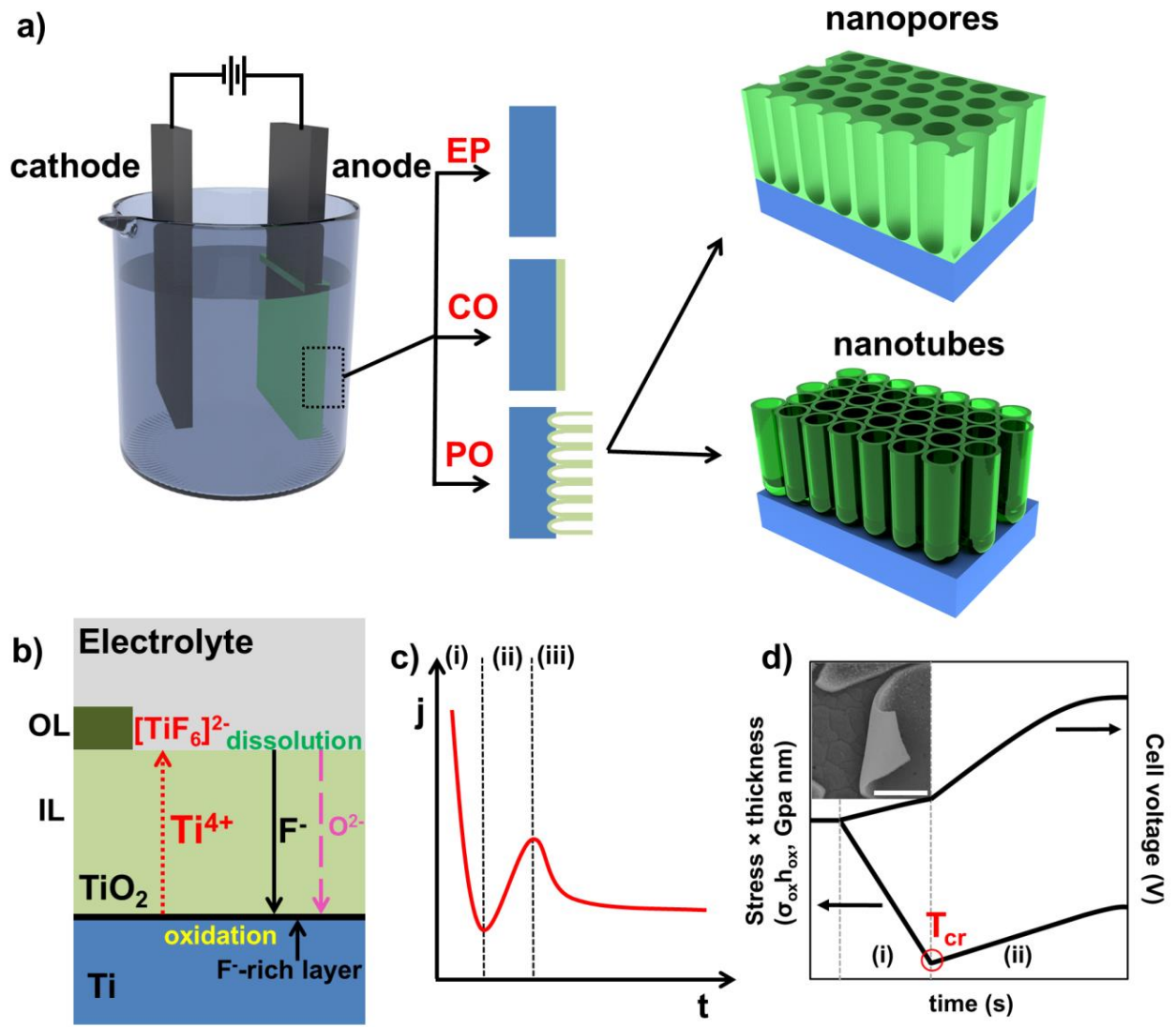

Figure 2

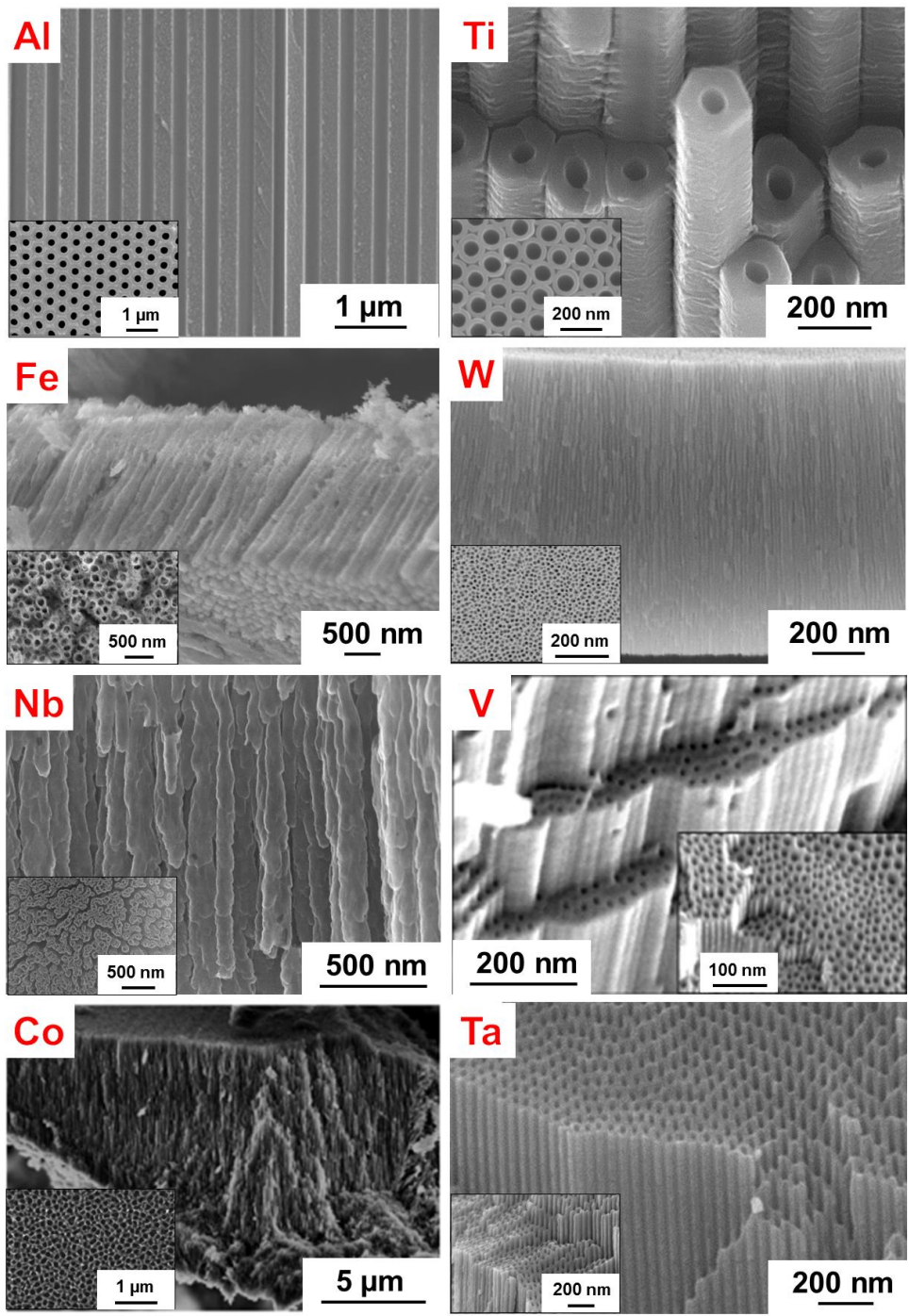

Figure 3

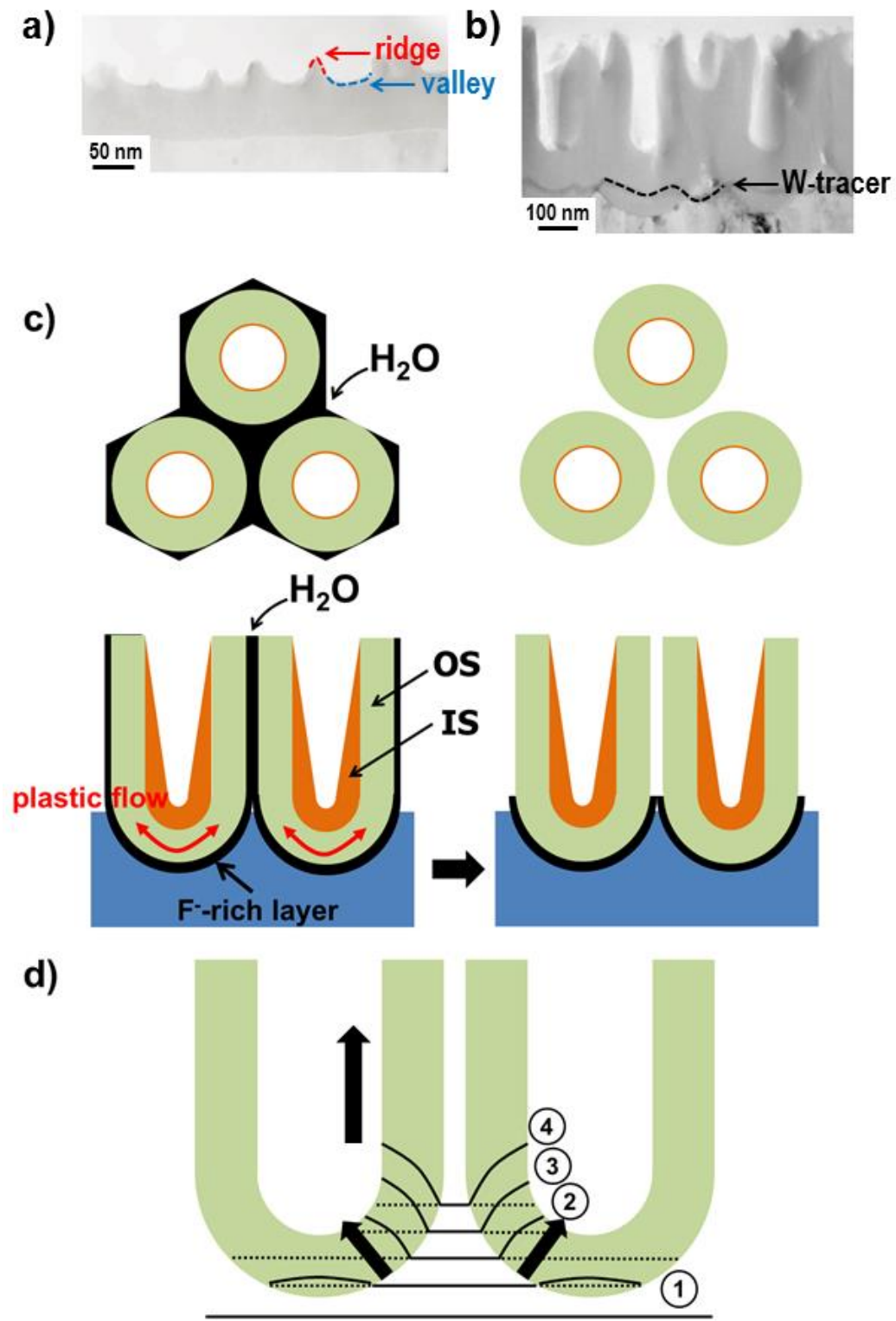

Figure 4

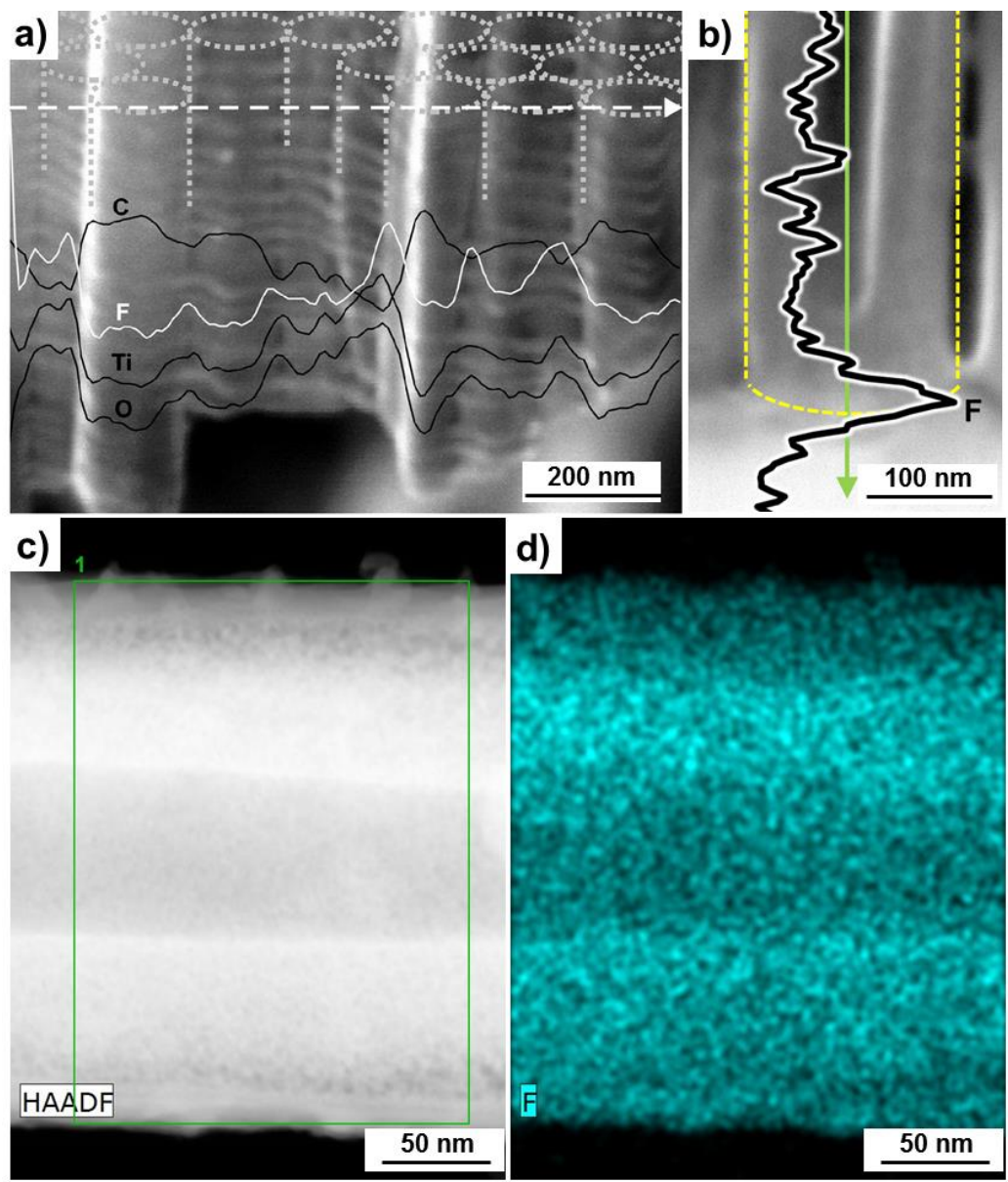

Figure 5

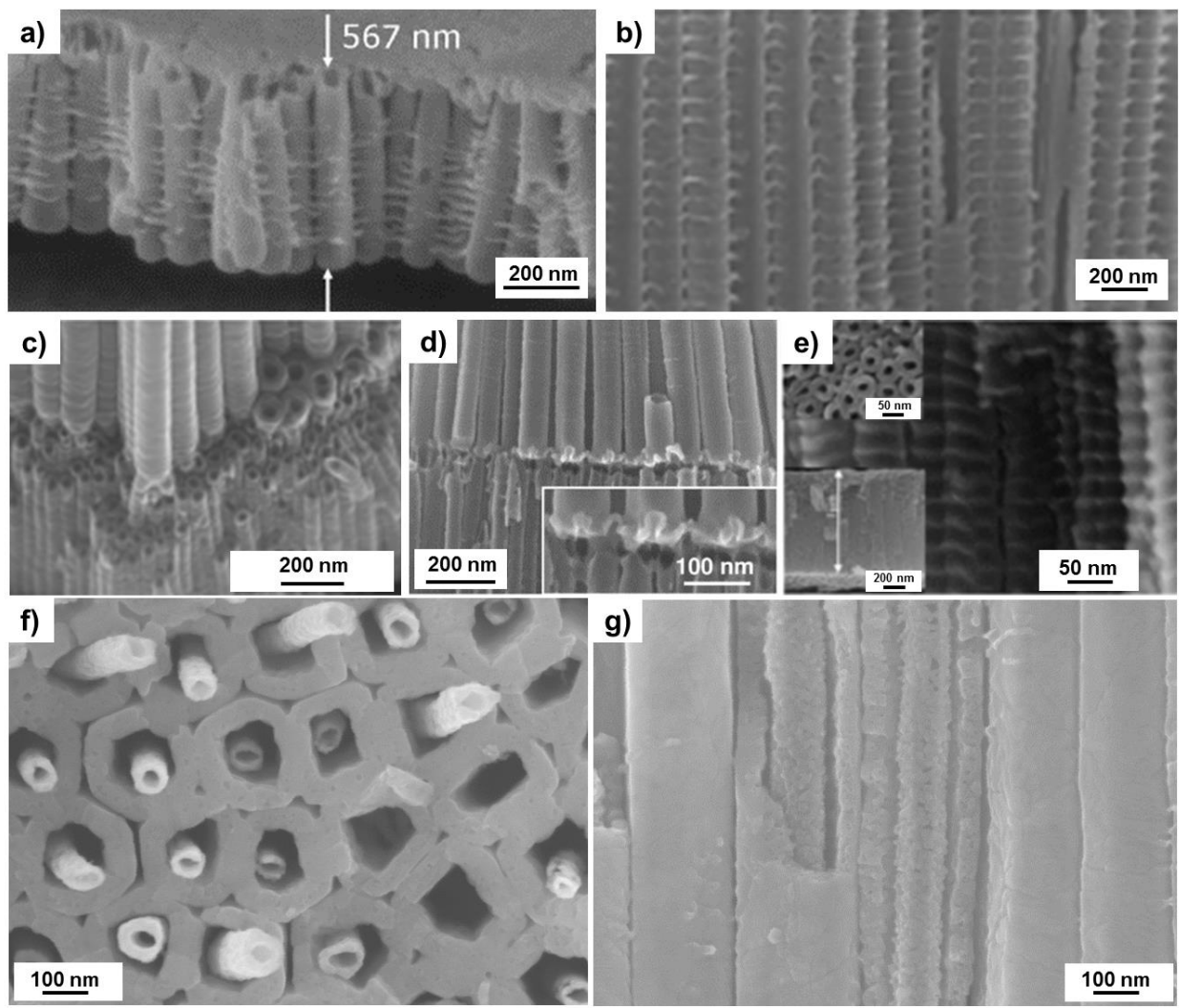

Figure 6

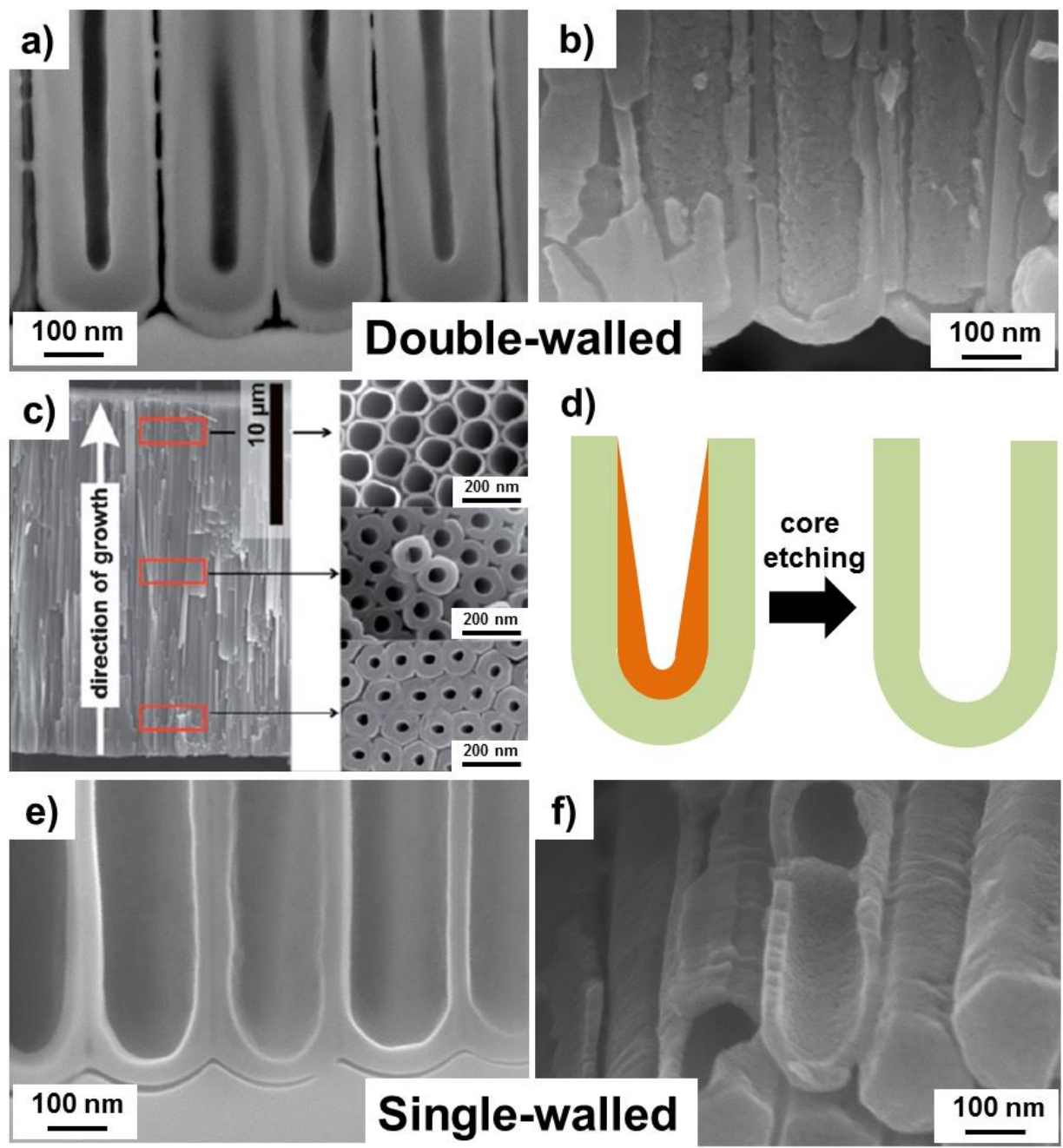

Figure 7

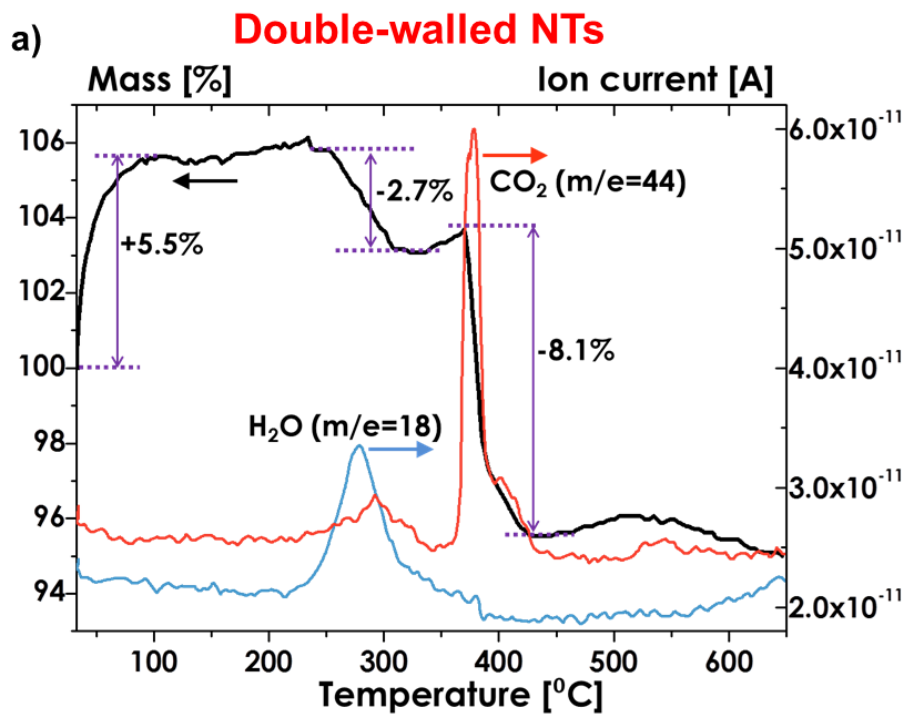

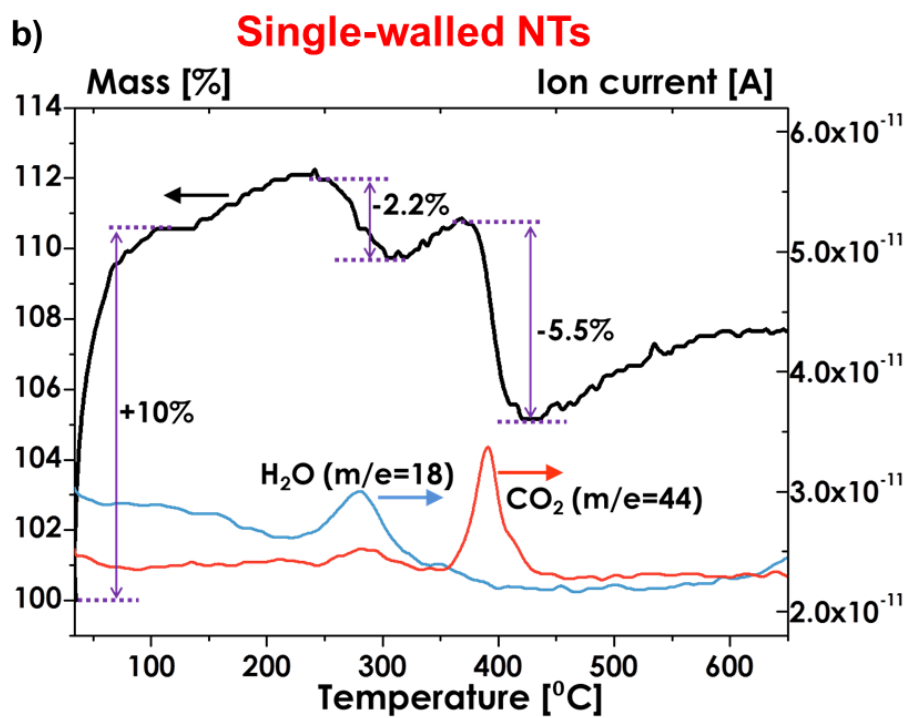

Figure 8

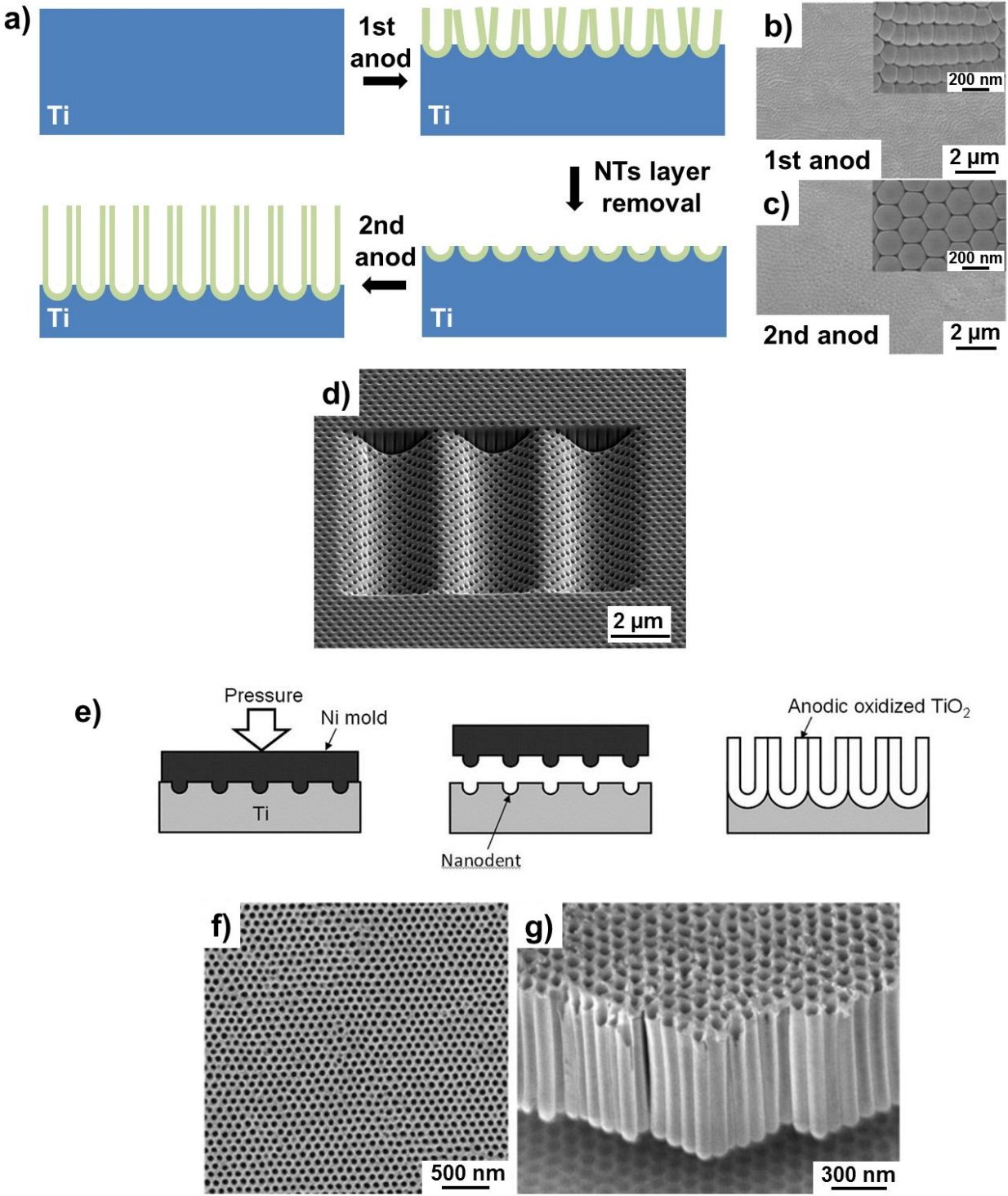

Figure 9

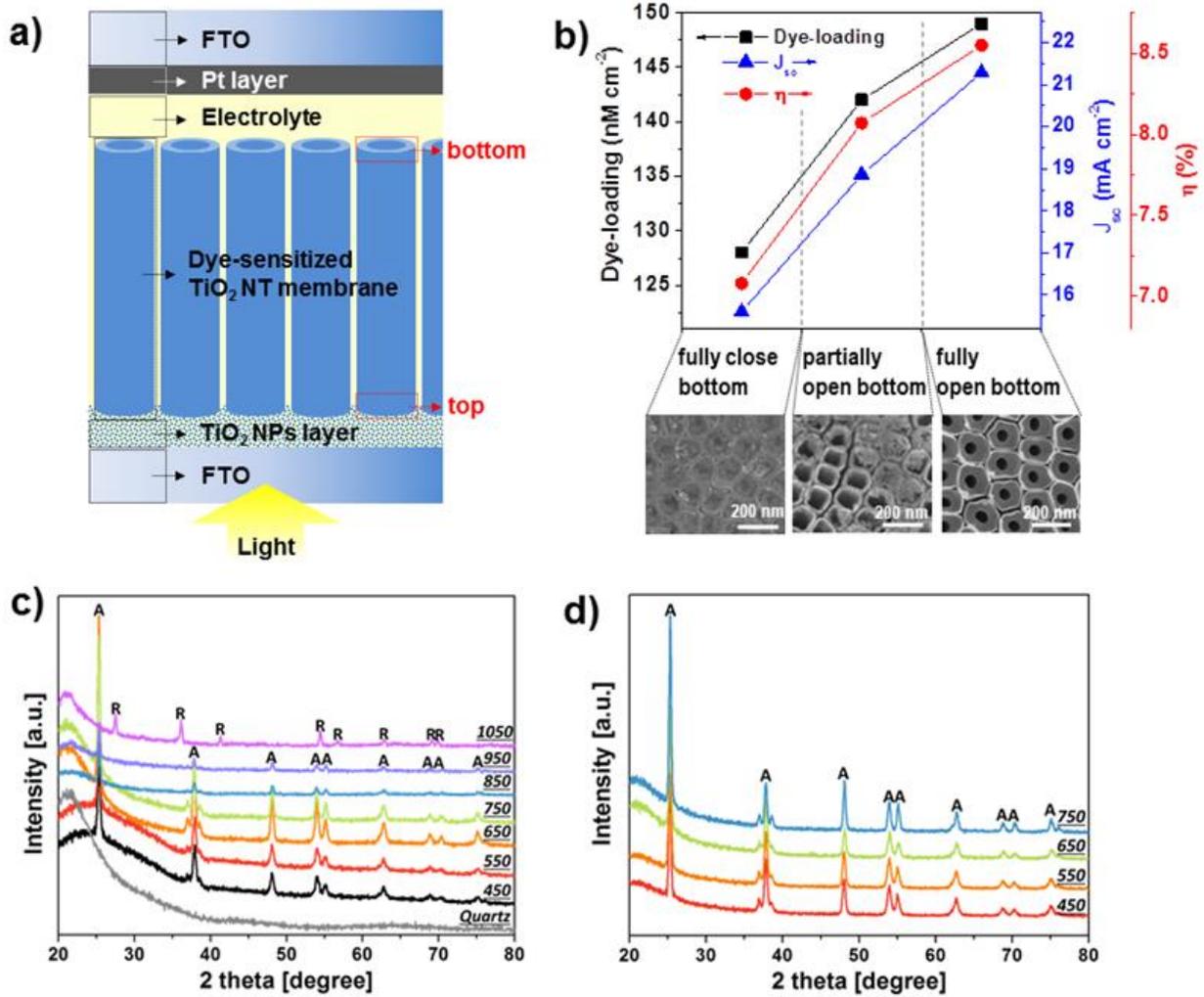

Figure 10

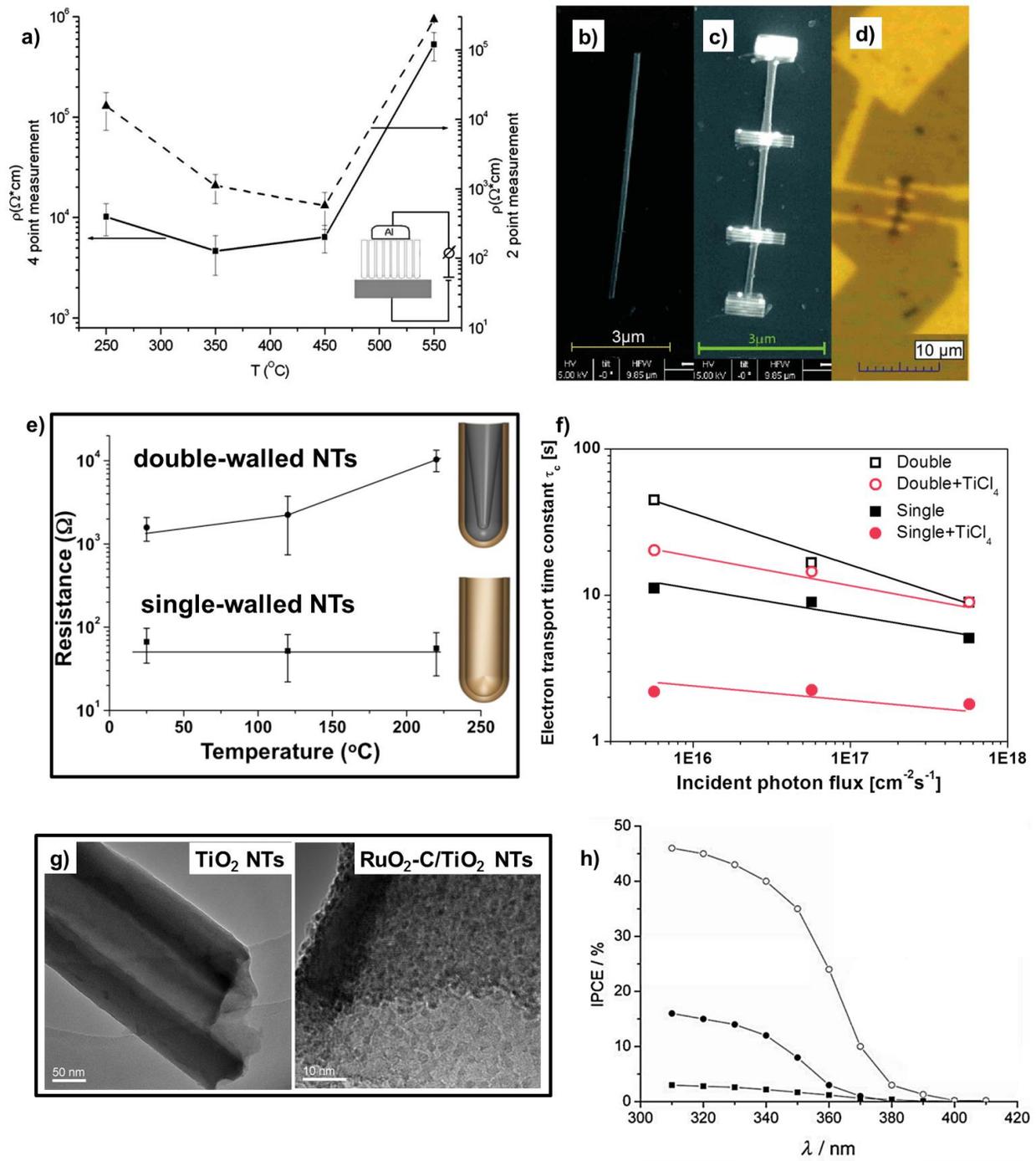